\documentclass[12pt,epsf]{article}
\usepackage{amscd,amssymb,epsf}

\advance \topmargin by -\headheight
\advance \topmargin by -\headsep

\evensidemargin \oddsidemargin
\def\ie{{\em i.e.}}
\def\ie{\hbox{\it i.e.}}

\def\CC{{\mathchoice
{\rm C\mkern-8mu\vrule height1.45ex depth-.05ex
width.05em\mkern9mu\kern-.05em}
{\rm C\mkern-8mu\vrule height1.45ex depth-.05ex
width.05em\mkern9mu\kern-.05em}
{\rm C\mkern-8mu\vrule height1ex depth-.07ex
width.035em\mkern9mu\kern-.035em}
{\rm C\mkern-8mu\vrule height.65ex depth-.1ex
width.025em\mkern8mu\kern-.025em}}}

\def\RR{{\rm I\kern-1.6pt {\rm R}}}

\def\ZZ{{\rm Z}\kern-3.8pt {\rm Z} \kern2pt}
\def\IB{\relax{\rm I\kern-.18em B}}
\def\ID{\relax{\rm I\kern-.18em D}}
\def\II{\relax{\rm I\kern-.18em I}}
\def\IP{\relax{\rm I\kern-.18em P}}

\def\np{Nucl. Phys.}

\def\pr{Phys. Rev.}

\def\jhep{J. High Energy Phys.}

\newcommand{\beq}{\begin{equation}}
\newcommand{\eeq}{\end{equation}}
\newcommand{\rc}{\nonumber\\}
\newcommand{\bear}{\begin{eqnarray}}
\newcommand{\eear}{\end{eqnarray}}

\def\to{\rightarrow}

\def\to{\rightarrow}

\newfont{\namefont}{cmr10}
\newfont{\addfont}{cmti7 scaled 1440}
\newfont{\boldmathfont}{cmbx10}
\newfont{\headfontb}{cmbx10 scaled 1728}
\def\ie{{\it i.e.}}

\def\revise#1       {\raisebox{-0em}{\rule{3pt}{1em}}%
                     \marginpar{\raisebox{.5em}{\vrule width3pt\
                     \vrule width0pt height 0pt depth0.5em
                     \hbox to 0cm{\hspace{0cm}{%
                     \parbox[t]{4em}{\raggedright\footnotesize{#1}}}\hss}}}}

\def\calq         {{\cal Q}}

\def\ee           {{\rm e}}

\def\sqr#1#2{{\vcenter{\vbox{\hrule height.#2pt
 \hbox{\vrule width.#2pt height#1pt \kern#1pt
 \vrule width.#2pt}\hrule height.#2pt}}}}

\def\a{\alpha}
\def\b{\beta}

\def\be{\begin{equation}}
\def\ee{\end{equation}}

\def\m{\mu}
\def\g{\gamma}

\def \ov{\over}

\def\ep{\epsilon}

\def \p {\phi}
\def \te {\tilde \epsilon}

\newcommand{\eel}[1]{\label{#1}\end{equation}}
\newcommand{\bea}{\begin{eqnarray}}
\newcommand{\eea}{\end{eqnarray}}
\newcommand{\eeal}[1]{\label{#1}\end{eqnarray}}

\def \ov {\over}

\def\ep {\epsilon}

\def \p {\phi}
\def \te {\tilde \epsilon}

\def\m{\mu}

\def\a{\alpha}
\def\b{\beta}
\def\g{\gamma}
\def\G{\Gamma}

\def\te{\theta}

\def\p{\phi}

\def\ep{\epsilon}

\def\be{\begin{equation}}
\def\ee{\end{equation}}

\def \g {\gamma}
\def \G {\Gamma}
\def \k {\kappa}

\def \m {\mu}

\def\te{\theta}
\def\g{\gamma}

\begin{document}
\begin{titlepage}

\begin{center} \Large \bf Supersymmetric Branes on ${\bf AdS_5 \times
Y^{p,q}}$ and their Field Theory Duals
\end{center}

\vskip 0.3truein
\begin{center}
Felipe Canoura${}^{\,*}$
\footnote{canoura@fpaxp1.usc.es},
Jos\'e D. Edelstein${}^{\,*\dagger}$
\footnote{edels@fpaxp1.usc.es},
Leopoldo A. Pando Zayas${}^{\,\ddagger}$
\footnote{lpandoz@umich.edu},\\
Alfonso V. Ramallo${}^{\,*}$
\footnote{alfonso@fpaxp1.usc.es}
and
Diana Vaman${}^{\,\ddagger}$
\footnote{diana@neu.physics.lsa.umich.edu}

\vspace{0.3in}

${}^{\,*}$Departamento de F\'\i sica de Part\'\i culas, Universidade de
Santiago de Compostela \\and
Instituto Galego de F\'\i sica de Altas Enerx\'\i as (IGFAE)\\
E-15782 Santiago de Compostela, Spain
\vspace{0.2in}

${}^{\,\dagger}$Centro de Estudios Cient\'\i ficos (CECS) Casilla 1469,
Valdivia, Chile
\vspace{0.2in}

${}^{\,\ddagger}$ Michigan Center for Theoretical Physics\\
Randall Laboratory of Physics, The University of Michigan\\
Ann Arbor, MI 48109-1040, USA

\end{center}
\vskip.5truein

\begin{center}
\bf ABSTRACT
\end{center}
We systematically study supersymmetric embeddings of D-brane probes of
different dimensionality in the $AdS_5\times Y^{p,q}$ background of type
IIB string theory. The main technique employed is the kappa symmetry of
the probe's worldvolume theory. In the case of D3-branes, we recover the
known three-cycles dual to the dibaryonic operators of the gauge theory
and we also find a new family of supersymmetric embeddings. The BPS
fluctuations of dibaryons are analyzed and shown to match the gauge
theory results. Supersymmetric configurations of D5-branes, representing
domain walls, and of spacetime filling D7-branes (which can be used to add
flavor) are also found. We also study the baryon vertex and some other
embeddings which break supersymmetry but are nevertheless stable. 

\vskip1.3truecm
\leftline{US-FT-3/05}
\leftline{CECS-PHY-05/15}
\leftline{MCTP-05-99}
\leftline{NSF-KITP-05-107}
\leftline{hep-th/0512087 \hfill December 2005}
\smallskip
\end{titlepage}
\setcounter{footnote}{0}


\setcounter{equation}{0}
\section{Introduction}
\medskip

The Maldacena conjecture provides a unique window into the strongly
coupled physics of gauge theories in terms of a string theory 
\cite{juan,MAGOO}. A crucial ingredient in the AdS/CFT correspondence is 
the state/operator correspondence. It provides the basis for 
explicit computations. Calculationally, it is convenient to consider the 
limit of large 't Hooft coupling where the supergravity 
approximation is valid. More precisely,  chiral operators of the CFT 
are in correspondence with the modes of supergravity in 
the dual background.  However, much information is contained in 
the stringy sector of the correspondence and some of it crucially
survives the large 't Hooft limit. 

For example, as discovered by Witten \cite{ba0} in discussing the
duality in the case of ${\cal N}=4$ super Yang--Mills (SYM) theory with
gauge group $SO(2N)$ and its dual $AdS_5\times \mathbb{RP}^5$, the gravity
side must contain branes, just to accommodate a class of chiral operators of
the gauge theory. The study of branes wrapped in the gravity theory becomes
an intrinsic part of the correspondence. It has been extended and understood
in a variety of situations. For example, a vertex connecting N fundamental strings --known as the baryon vertex-- can be identified with a baryon built
out of external quarks, since each string ends on a charge in the fundamental
representation of $SU(N)$. Such an object can be constructed by wrapping a
D5-brane over the whole five-dimensional compact manifold \cite{ba0}.
Also, domain walls in the field theory side can be understood as D5-branes
wrapping 2-cycles of the internal geometry \cite{ba0,ba1}. In quantum
field theories that arise in D3-branes placed at conical singularities,
an object of particular interest is given by D3-branes wrapped on
supersymmetric 3-cycles; these states are dual to dibaryons built
from chiral fields charged under two different gauge groups of the resulting
quiver theory \cite{ba0,ba1,ba2,ba3}. In the absence of a string theory
formulation on backgrounds with Ramond-Ramond forms, probe D-branes of
various dimensions provide valuable information about the spectrum.
More generally, finding particular situations where a semiclassical
description captures nontrivial stringy information is an important
theme of the AdS/CFT correspondence recently fueled largely by BMN in
\cite{BMN}, but having its root in the work of Witten \cite{ba0} and in
considerations of the Wilson loop as a classical string in the
supergravity background \cite{wl}. 

Given a Sasaki-Einstein five-dimensional manifold $X^5$ one can consider
placing a stack of $N$ D3-branes at the tip of the (Calabi--Yau) cone over
$X^5$. Taking the Maldacena limit then leads to a duality between string
theory on $AdS_5\times X^5$ and a superconformal gauge theory living in the
worldvolume of the D3-branes \cite{gubser}. When the Sasaki--Einstein
manifold is the $T^{1,1}$ space --the Calabi--Yau cone being nothing but
the conifold-- we have the so-called Klebanov--Witten
model \cite{kw}, which is dual to a four-dimensional ${\cal N}=1$
superconformal field theory with gauge group $SU(N)\times SU(N)$ coupled
to four chiral superfields in the bifundamental representation. Important
aspects of this duality, relevant in the context of this article, have been
further developed in \cite{ba1,ba2,beasley}. The supersymmetry of
D-brane probes in the Klebanov--Witten model was studied in full detail
in ref.\cite{acr}. 

Recently, a new class of Sasaki-Einstein manifolds $Y^{p,q}$, $p$ and $q$
being two coprime positive integers, has been constructed \cite{GMSW1,GMSW2}.
The infinite family of spaces $Y^{p,q}$ was shown to be dual to superconformal
quiver gauge theories \cite{ms,sequiver}. The study of AdS/CFT in these
geometries has shed light in many subtle aspects of superconformal field
theories in four dimensions. Furthermore, the correspondence successfully
passed new tests such as those related to the fact that the central charge
of these theories, as well as the R-charges of the fundamental fields, are
irrational numbers \cite{friends}.

In this paper we perform a systematic classification of supersymmetric
branes in  the $AdS_5\times Y^{p,q}$ geometry and study their field
theoretical interpretation. It is worth reminding that
the spectrum of IIB supergravity compactified on $Y^{p,q}$ is not known
due to various technical difficulties including the general form of Heun's
equation. Therefore, leaving aside the chiral primaries, very little is
known about the gravity modes dual to protected operators in the field
theory. Our study of supersymmetric objects in the gravity side is a way
to obtain information about properties of these operators in the gauge
theory side. They comprise interesting physical objects of these theories
such as the baryon vertex, domain walls, the introduction of flavor, fat
strings, etc. It is very remarkable that we are able to provide precise
information about operators with large conformal dimension that grows
like $N$. Moreover, we can also extract information about excitations of
these operators.

The main technique we employ to determine the supersymmetric embeddings
of D-brane probes in the $AdS_5\times Y^{p,q}$ background is
kappa symmetry \cite{swedes} and follows the same systematics as in the
analysis performed in ref.\cite{acr} for the case of the  $AdS_5\times
T^{1,1}$ background. Our approach is based on the existence of a matrix
$\Gamma_{\kappa}$ which depends on the metric induced on the worldvolume
of the probe and characterizes its supersymmetric embeddings. Actually, if
$\epsilon$ is a Killing spinor of the background, only those embeddings
such that $\Gamma_{\kappa}\,\epsilon\,=\,\epsilon$ preserve some supersymmetry
\cite{bbs}. This kappa symmetry condition gives rise to a set of
first-order BPS differential equations whose solutions, if they exist,
determine the embedding
of the probe and the fraction of the original background supersymmetry that
it preserves. The configurations found by solving these equations also solve
the equations of motion derived from the Dirac-Born-Infeld action of the
probe and, actually, we will verify that they saturate a bound for the energy,
as it usually happens in the case of worldvolume solitons \cite{GGT}. 

The first case we study is that of D3-branes. We are able to find in this
case the three-cycles introduced in refs. \cite{ms,sequiver,hek} to
describe the different dibaryonic operators of the gauge theory.
Moreover, we also find a general class of supersymmetric embeddings of
the D3-brane probe characterized by a certain local holomorphicity
condition. Contrary to what happens in the case of $T^{1,1}$, globally it
turns out that these embeddings, in general, do not define a three-cycle
but a submanifold with boundaries. We also study the fluctuations of the
D3-brane probe around a dibaryonic configuration and we  successfully
match the emerging results with those of the corresponding quiver theory.
We also find stable non-supersymmetric configurations of D3-branes
wrapping a two-dimensional submanifold of $Y^{p,q}$ that define a one
dimensional object in the gauge theory that could be interpreted as a
{\it fat} string.

Our analysis continues with the study of D5-brane probes. We find
embeddings in which the D5-brane wraps a two-dimensional submanifold and
creates a domain wall in $AdS_5$. When crossing these domain walls the
rank of the gauge group factors of the quiver gets shifted \cite{ba1},
this leading to their identification as fractional branes \cite{klenek}.
We also study other stable configurations that break supersymmetry
completely but are nevertheless interesting enough. One of these
configurations is the baryon vertex, in which the D5-brane wraps the
entire $Y^{p,q}$ space.  Besides, we also consider the case of D5-branes
wrapping a two-dimensional submanifold when a nonvanishing worldvolume
flux is present as well as the setting with D5-branes on three-cycles of
$Y^{p,q}$.

Finally we turn to the case of D7-brane probes. According to the original
proposal of ref.\cite{karch}, the embeddings in which the D7-branes fill
completely the gauge theory directions are specially interesting. These
spacetime filling configurations can be used as flavor branes, \ie\ as
branes whose fluctuations can be identified with the dynamical mesons of
the gauge theory (see refs.\cite{mesons}-\cite{Decays} for the
analysis of the meson spectrum in different theories). Moreover, we
show that the configurations in which the D7-brane wraps the entire
$Y^{p,q}$ are also supersymmetric. 

The organization of the paper is as follows. In section \ref{ypq} we review
some properties of the $Y^{p,q}$ space and the corresponding Calabi--Yau
cone that we call $CY^{p,q}$, including the local complex coordinates of
the latter. We discuss the corresponding type IIB supergravity solution
and present the explicit form of its Killing spinors. We also present the
general condition satisfied by supersymmetric embeddings of D-brane probes
on this background. Section \ref{d3} discusses embeddings of D3-branes on
various supersymmetric cycles. We reproduce the three-cycles considered
previously in the literature and  find a new family of supersymmetric
embeddings. Section \ref{d3} also contains an analysis of the excitations
of wrapped D3-branes and we find perfect agreement with the corresponding
field theory results. Section \ref{d5} deals with supersymmetric
D5-branes which behave as domain walls, while in section \ref{d7} we
discuss the spacetime filling embeddings of D7-branes. For completeness,
we consider other possible embeddings, such as the baryon vertex,  in
section \ref{6}. We conclude and summarize our results in section
\ref{conclusions}.

\setcounter{equation}{0}
\section{The $Y^{p,q}$ space and brane probes}
\medskip
\label{ypq}

Let us consider a solution of IIB supergravity given by a ten-dimensional
space whose metric is of the form:
\beq
ds^2\,=\,ds^2_{AdS_5}\,+\,L^2\,ds^2_{Y^{p,q}}
\label{10dmetric}
\eeq
where $ds^2_{AdS_5}$ is the metric of $AdS_5$ with radius $L$
and $ds^2_{Y^{p,q}}$ is the metric of the Sasaki-Einstein space $Y^{p,q}$,
which can be written as \cite{GMSW1, GMSW2}:
\bear
&&ds^2_{Y^{p,q}}\,=\,{1-cy\over
6}\,(d\theta^2\,+\,\sin^2\theta\,d\phi^2)\,+\, {1\over
6\,H^2(y)}\,dy^2\,+\,{H^2(y)\over 6}\, 
(d\beta\,-\,c\cos\theta d\phi)^2\,\rc\rc
&&\qquad\qquad+\,{1\over 9}\,
\big[\,d\psi\,+\,\cos\theta d\phi\,+\,y(d\beta\,-\,c\cos\theta d\phi)\,
\big]^2\,\,,
\label{Ypqmetric}
\eear
$H(y)$ being given by:
\beq
H(y)=\sqrt{{a-3y^2+2cy^3\over 3(1-cy)}}\,\,.
\label{Hfunction}
\eeq
The metrics $ds^2_{Y^{p,q}}$ are Sasaki-Einstein, which means that the
cones $CY^{p,q}$ with metric $dr^2+r^2 ds^2_{Y^{p,q}}$ are Calabi-Yau
manifolds. The metrics in these coordinates neatly display some nice local
features of these spaces. Namely, by writing it as
\beq
ds^2_{Y^{p,q}}\,=\,ds^2_{4} + \left[ \frac{1}{3} d\psi + \sigma
\right]^2 ~,
\label{slice}
\eeq
it turns out that $ds^2_{4}$ is a K\"ahler--Einstein metric with K\"ahler
form $J_4 = \frac{1}{2} d\sigma$. Notice that this is a local splitting
that carries no global information. Indeed, the pair $(ds^2_{4},J_4)$ is
not in general globally defined. The Killing vector $\frac{\partial
~}{\partial\psi}$ has constant norm but its orbits do not close (except
for certain values of $p$ and $q$, see below). It defines a foliation
of $Y^{p,q}$ whose transverse leaves, as we see, locally have a
K\"ahler--Einstein structure. This aspect will be important in later
discussions.

These $Y^{p,q}$ manifolds are topologically $S^2\times S^3$ and can be
regarded as $U(1)$ bundles over manifolds of topology  $S^2\times S^2$.
Its isometry group is $SU(2) \times U(1)^2$. Notice that the metric
(\ref{Ypqmetric}) depends on  two constants $a$ and $c$. The latter,
if different from zero, can be set to one by a suitable rescaling of the
coordinate $y$, although it is sometimes convenient to keep the value of
$c$ arbitrary in order to be able to recover the $T^{1,1}$ geometry, which
corresponds to $c=0$ \footnote{If $c=0$, we can set $a = 3$ by rescaling
$y \to \xi y$, $\beta \to \xi^{-1} \beta$, and $a \to \xi^2 a$. If we
further write $y = \cos\tilde\theta$ and $\beta = \tilde\phi$, and
choose the period of $\psi$ to be $4\pi$, the metric goes to that of
$T^{1,1}$.}. If $c\neq 0$, instead, as we have just said we can set 
 $c = 1$ and the parameter $a$ can be written in terms of two
coprime integers $p$ and $q$ (we take $p>q$) as follows:
\beq
a\,=\,{1\over 2}\,-\,{p^2-3q^2\over 4p^3}\,\,\sqrt{4p^2-3q^2}\,\,.
\eeq
Moreover, the coordinate $y$ ranges between the two smaller roots of the
cubic equation
\beq
\calq(y) \equiv a - 3y^2 + 2cy^3\,=\,2c\,\prod_{i=1}^{3}\,
(y-y_i) ~,
\label{calq}
\eeq
\ie\ $\,y_1 \le y \le y_2$ with (for $c=1$):
\bear
&&y_1\,=\,{1\over 4p}\,
\Big(\,2p\,-\,3q\,-\,\sqrt{4p^2-3q^2}\,\Big)\,< 0\,,\rc
&&y_2\,=\,{1\over 4p}\,
\Big(\,2p\,+\,3q\,-\,\sqrt{4p^2-3q^2}\,\Big)\,> 0\,.
\label{y12}
\eear
In order to specify the range of the other variables appearing in the metric,
let us introduce the coordinate $\alpha$ by means of the relation:
\beq
\beta\,=\,-(6\alpha+c\psi)\,\,.
\label{beta-alpha}
\eeq
Then, the coordinates $\theta$,$\phi$,$\psi$ and $\alpha$ span the range:
\beq
0\le \theta\le \pi\,\,,\qquad
0< \phi\le 2\pi\,\,,\qquad
0< \psi\le 2\pi\,\,,\qquad
0< \alpha\le 2\pi \ell\,\,,
\eeq
where $\ell$ is (generically an irrational number) given by:
\beq
\ell\,=\,-{q\over 4p^2\,y_1\,y_2}\,=\,{q\over 3q^2\,-\,2p^2\,+\,p
\sqrt{4p^2\,-\,3q^2}}\,\,,
\label{alphaperiod}
\eeq
the metric (\ref{Ypqmetric}) being periodic in these variables. Notice that,
whenever $c\neq 0$, the coordinate $\beta$ is non-periodic: the periodicities
of $\psi$ and $\alpha$ are not congruent, unless the manifold is quasi-regular,
{\it i.e.}, there exists a positive integer $k$ such that
\beq
k^2 = 4 p^2 - 3 q^2 ~.
\label{qreg}
\eeq
For quasi-regular manifolds, $ds^2_{4}$ in (\ref{slice}) corresponds to a
K\"ahler--Einstein orbifold. Notice that $\ell$ becomes rational and it is
now possible to assign a periodicity to $\psi$ such that $\beta$ ends up
being periodic. If we perform the change of variables (\ref{beta-alpha})
in (\ref{Ypqmetric}), we get
\bear 
&&ds^2_{Y^{p,q}}\,=\,{1-cy\over
6}\,(d\theta^2\,+\,\sin^2\theta\,d\phi^2)\,+\, {1\over
6\,H^2(y)}\,dy^2\,+\,{v(y)\over 9}\, 
(d\psi\,+\,\cos\theta d\phi)^2\,+\,\rc\rc
&&\qquad\qquad+\,w(y)\,
\big[\,d\alpha\,+\,f(y) \left( d\psi
+ \cos\theta d\phi \right) \big]^2\,\,,
\label{alphametric}
\eear
with $v(y)$, $w(y)$ and $f(y)$ given by
\beq
v(y) = {a - 3y^2 + 2cy^3 \over a - y^2} ~, ~~~~~
w(y) = {2 (a - y^2) \over 1 - cy} ~, ~~~~~
f(y) = {ac - 2y + y^2c \over 6 (a - y^2)}~.
\label{qwy}
\eeq

Concerning the $AdS_5$ space, we will represent it by means of four Minkowski
coordinates $x^{\alpha}$ ($\alpha=0,1,2,3$) and a radial variable $r$. In
these coordinates, the $AdS_5$ metric takes the standard form:
\beq
ds^2_{AdS_5}\,=\,{r^2\over L^2}\,dx^2_{1,3}\,+\,{L^2\over r^2}\,dr^2\,\,.
\label{AdSmetric}
\eeq
The ten-dimensional metric (\ref{10dmetric}) is then a solution of the
equations of motion of type IIB supergravity if, in addition, we have $N$
units of flux of the self-dual Ramond-Ramond five-form $F^{(5)}$. This
solution corresponds to the near-horizon region of a stack of $N$ coincident
D3-branes extended along the Minkowski coordinates and located at the apex
of the $CY^{p,q}$ cone. The explicit expression of $F^{(5)}$ is:
\beq
g_s\,F^{(5)}=d^4x\,\wedge dh^{-1}\,+\,{\rm Hodge\,\,\, dual}\,\,,
\label{F5}
\eeq
where $h(r)$ is the near-horizon harmonic function, namely:
\beq
h(r)={L^4\over r^4}\,\,.
\eeq
The quantization condition of the flux of $F^{(5)}$ determines the constant
$L$ in terms of $g_s$, $N$, $\alpha'$ and the volume ${\rm Vol}(Y^{p,q})$
of the Sasaki-Einstein space:
\beq
L^4\,=\,{4\pi^4\over {\rm Vol}(Y^{p,q})}\,g_s\,N\,(\alpha')^2\,\,,
\label{L}
\eeq
where ${\rm Vol}(Y^{p,q})$ can be computed straightforwardly from the metric
(\ref{Ypqmetric}), with the result (for $c=1$):
\beq
{\rm Vol}(Y^{p,q})\,=\,{q^2\over 3p^2}\,\,
{2p+\sqrt{4p^2-3q^2}\over 3q^2-2p^2+p\sqrt{4p^2-3q^2}}\,\,\pi^3\,\,.
\eeq

\subsection{Quiver theories for $Y^{p,q}$ spaces}
The gauge theory dual to IIB on $AdS_5\times Y^{p,q}$ is by now well
understood. Here we quote some of the results that are directly relevant to
our discussion. We follow the presentation of ref.~\cite{sequiver}.

\begin{figure}[ht]
\centerline{\epsffile{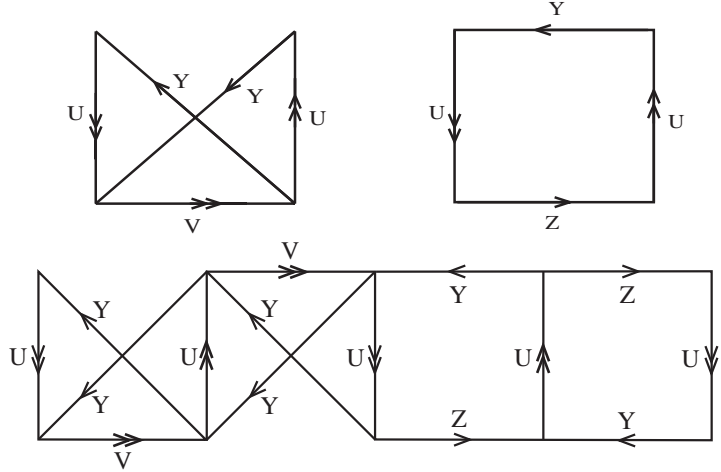}}
\caption{The basic cells $\sigma$ (upper left) and $\tau$ (upper right).
$Y^{p,q}$ quivers are built with $q$ $\sigma$ and $p-q$ $\tau$ unit
cells. The cubic terms in the superpotential (\ref{supYpq}) come from
closed loops of the former and the quartic term arises from the latter.
The quiver for $Y^{4,2}$ is given by $\sigma\tilde\sigma\tau\tilde\tau$
(bottom).}
\label{Y42}
\end{figure}

The quivers for $Y^{p,q}$ can be constructed starting with the quiver
of $Y^{p,p}$ which is naturally related to the quiver theory obtained
from $\mathbb C^3/\mathbb Z_{2p}$. The gauge group is $SU(N)^{2p}$ and the
superpotential is constructed out of cubic and quartic terms in the
four types of bifundamental chiral fields present: two doublets $U^\a$
and $V^\b$ and two singlets $Y$ and $Z$ of a global $SU(2)$. Namely,
\beq
W=\sum\limits_{i=1}^q \epsilon_{\a\b} (U^\a_i V^\b_i Y_{2i-1}+ V^\a_i U^\b_{i+1}Y_{2i})
+ \sum\limits_{j=q+1}^p \epsilon_{\a\b}Z_j U_{j+1}^\a Y_{2j-1} U_{j}^\b.
\label{supYpq}
\eeq
Greek indices $\a,\b=1,2$ are in $SU(2)$, and Latin subindices $i,j$ refer
to the gauge group where the corresponding arrow originates. Equivalently,
as explained in \cite{hek}, the quiver theory for $Y^{p,q}$ can be
constructed from two basic cells denoted by $\sigma$ and $\tau$, and their
mirror images with respect to a horizontal axis, $\tilde\sigma$ and
$\tilde\tau$ (see Fig.\ref{Y42}). Gluing of cells has to respect the
orientation of double arrow lines corresponding to the $U$ fields. For
example, the quiver $Y^{4,2}$ is given by
$\sigma\tilde\sigma\tau\tilde\tau$. More concrete examples and further
discussion can be found in \cite{sequiver,hek}.


Here we quote a result of \cite{sequiver} which we will largely reproduce
using a study of wrapped branes. The global $U(1)$ symmetries corresponding
to the factors appearing in the isometry group of the $Y^{p,q}$ manifold are
identified as the R-charge symmetry $U(1)_R$ and a flavor symmetry $U(1)_F$.
There is also a baryonic $U(1)_B$ that becomes a gauge symmetry in the
gravity dual. The charges of all fields in the quiver with respect to
these Abelian symmetries is summarized in Table \ref{charges}.

\begin{table}[ht]
\begin{center}
$$\begin{array}{|c|c|c|c|c|}
\hline
\mathrm{Field}&\mathrm{number}&R -\mathrm{charge}& U(1)_B &  U(1)_F \\
\hline\hline
& & & & \\[-1ex]
Y      & p + q
	& {- 4 p^2 + 3 q^2 + 2 p q + (2 p - q)\sqrt{4 p^2 - 3 q^2} \ov 3 q^2}
	& p - q &- 1 \\[1.5ex]
\hline
& & & & \\[-1ex]
Z      & p - q
	& {- 4 p^2 + 3 q^2 - 2 p q + (2 p + q)\sqrt{4 p^2 - 3 q^2} \ov 3 q^2}
	& p + q &+ 1 \\[1.5ex]
\hline
& & & & \\[-1ex]
U^{\alpha} &  p
	& {2 p (2 p -  \sqrt{4 p^2 - 3 q^2}) \ov 3 q^2} & - p & 0  \\[1.5ex]
\hline
& & & & \\[-1ex]
V^{\beta} &  q
	& {3 q - 2 p + \sqrt{4 p^2 - 3 q^2} \ov 3 q} & q & + 1 \\[1.5ex]
\hline
\end{array}$$
\caption{Charges for bifundamental chiral fields in the quiver dual to
$Y^{p,q}$ \cite{sequiver}.}
\label{charges}
\end{center}
\end{table}
It is worth noting that the above assignment of charges satisfies a number
of conditions. For example, the linear anomalies vanish
$\mathrm{Tr}\,U(1)_B = \mathrm{Tr}\,U(1)_F = 0$, as well as the cubic
t'~Hooft anomaly
$\mathrm{Tr}\,U(1)_B^3$.

\subsection{Complex coordinates on $CY^{p,q}$}

We expect some of the supersymmetric embeddings of probes that will be
studied in the present paper to be related to the complex coordinates
describing $CY^{p,q}$. The relevant coordinates were introduced in \cite{ms}.
(Here we follow the notation of \cite{ben}.) The starting point in
identifying a good set of complex coordinates is the following set of
closed one-forms \cite{ms}:
\begin{eqnarray}
{\eta}^1 & = & \frac{1}{\sin\theta}\, d\theta - i d\phi ~, \nonumber\\
\tilde{\eta}^2 & = & -\frac{dy}{H(y)^2} - i (d\beta - c\cos\theta
d\phi) ~, \nonumber\\
\tilde{\eta}^3 & = & 3 \frac{dr}{r} + i \big[ d\psi + \cos\theta d\phi
+ y (d\beta - c\cos\theta d\phi) \big] ~,
\end{eqnarray}
in terms of which, the metric of $CY^{p,q}$ can be rewritten as
\begin{equation}
ds^2 = r^2 \frac{(1-cy)}{6}\, {\sin}^2\theta ~|{\eta}^1|^2 + r^2
\frac{H(y)^2}{6} ~|{\tilde{\eta}^2}|^2 + \frac{r^2}{9}
~|{\tilde{\eta}^3}|^2 ~.
\end{equation}
Unfortunately, ${\tilde{\eta}^2}$ and ${\tilde{\eta}^3}$ are not integrable.
It is however easy to see that integrable one-forms can be obtained by taking
linear combinations of them:
\be
{\eta}^2 = {\tilde{\eta}^2}+c \cos \theta ~{\eta}^1 ~, \qquad
{\eta}^3 = {\tilde{\eta}^3}+\cos\theta ~{\eta}^1 + y ~{\tilde{\eta}^2} ~.
\ee
We can now define ${\eta}^i=dz_i/z_i$ for $i=1,2,3$, where
\beq
z_1 = \tan\frac{\theta}{2}\, e^{-i\phi} ~, \qquad
z_2 = \frac{(\sin\theta)^{c}}{f_1(y)}\, e^{-i\beta} ~, \qquad
z_3 = r^3\, \frac{\sin\theta}{f_2(y)}\, e^{i\psi}\,\,,
\label{complexzs}
\eeq
with $f_1(y)$ and $f_2(y)$ being given by:
\begin{equation}
f_1(y)=\exp\left(\int\frac{1}{H(y)^2}dy\right),\qquad
f_2(y)=\exp\left(\int\frac{y}{H(y)^2}dy\right).
\label{fs}
\end{equation}
By using the form of $H(y)$ written in eq.(\ref{Hfunction}) it is possible
to provide a simpler expression for the functions $f_i(y)$, namely:
\bear
&&{1\over f_1(y)}\,=\,\sqrt{(y-y_1)^{{1\over y_1}}\,
(y_2-y)^{{1\over y_2}}\,(y_3-y)^{{1\over y_3}}}\,\,,\rc\rc
&&{1\over f_2(y)}\,=\,\sqrt{\calq(y)}\,=\,
\sqrt{2c}\,\sqrt{(y-y_1)\,(y_2-y)\,(y_3-y)}\,\,,
\label{yexpression}
\eear
where $\calq(y)$ has been defined in (\ref{calq}), $y_1$ and $y_2$ are
given in eq.(\ref{y12}) and $y_3$ is the third root of the polynomial
$\calq(y)$ which, for $c=1$, is related to $y_{1,2}$ as $y_3={3\over
2} - y_1 - y_2$. The holomorphic three-form of $CY^{p,q}$ simply reads  
\beq
\Omega = -\frac{1}{18} e^{i\psi} r^3 \sqrt{\frac{\calq(y)}{3}}
\sin\theta ~\eta^1 \wedge \eta^2 \wedge \eta^3 = -\frac{1}{18\sqrt{3}}
\frac{dz_1 \wedge dz_2 \wedge dz_3}{z_1 z_2} ~.
\label{threeform}
\eeq
Notice that coordinates $z_1$ and $z_2$ are local complex coordinates on the
transverse leaves of $Y^{p,q}$ (\ref{slice}) with K\"ahler--Einstein metric $ds_4^2$. They are not globally well-defined as soon as $z_2$ is periodic in
$\beta$ --which is not a periodic coordinate. Besides, they are meromorphic
functions on $CY^{p,q}$ (the function $z_1$ is singular at $\theta = \pi$
while $z_2$ has a singularity at  $y = y_1$). A set
of holomorphic coordinates on $Y^{p,q}$ was constructed in \cite{BHOP}.

\subsection{Killing spinors for $AdS_5\times Y^{p,q}$}

The $AdS_5\times Y^{p,q}$ background preserves eight supersymmetries, in
agreement with the ${\cal N}=1$ superconformal character of the corresponding
dual field theory, which has four ordinary supersymmetries and four
superconformal ones. In order to verify this statement, and for later use,
let us write explicitly the form of the Killing spinors of the background,
which are determined by imposing the vanishing of the supersymmetric
variations of the dilatino and gravitino. The result of this calculation
is greatly simplified in some particular basis of frame one-forms, which
we will now specify. In the $AdS_5$ part of the metric we will choose the
natural basis of vielbein one-forms, namely:
\beq
e^{x^{\alpha}}\,=\,{r\over L}\,\,dx^{\alpha}
\,\,,\,\,\,\,\,\,\,\, (\alpha=0,1,2,3)\,\,,
\,\,\,\,\,\,\,\,\,\,\,\,\,\,\,\,
e^{r}\,=\,{L\over r}\,\,.
\label{AdSvielbein}
\eeq
Moreover, in the $Y^{p,q}$ directions we will use the following frame:
\bea
e^1 &=& -\frac{L}{\sqrt{6}}\, \frac{1}{H(y)}\, dy ~, \nonumber \\
e^2 &=& -\frac{L}{\sqrt{6}}\, H(y)\, (d\beta- c\, \cos\theta\,
	d\phi) ~, \nonumber \\ 
e^3 &=& \frac{L}{\sqrt{6}}\, \sqrt{1-c\, y}\, d\theta\,\,, \nonumber \\
e^4 &=& \frac{L}{\sqrt{6}}\, \sqrt{1-c\,y}\sin\theta\, d\phi, \nonumber \\
e^5 &=& {L\over 3}\, \left( d\psi + y\, d\beta +(1-c\, y) \cos\theta\,
d\phi\right) ~.
\label{Ypqvielbein}
\eea
In order to write the expressions of the Killing spinors in a compact form, 
let us define the matrix $\Gamma_{*}$ as:
\beq
\Gamma_{*}\equiv i\Gamma_{x^0x^1x^2x^3}\,\,.
\label{gamma*}
\eeq
Then, the Killing spinors $\epsilon$ of the $AdS_5\times Y^{p,q}$ background 
can be written in terms of a constant spinor $\eta$ as \footnote{Note that
this spinor differs from the one of \cite{acr} by a rotation generated by
$e^{-{i\over 2}\psi\, \Gamma_{34}}$. This rotation accounts for the
difference between both frames.}:
\beq
\epsilon\,=\,e^{-{i\over 2}\psi}\, r^{-{\Gamma_{*}\over 2}}\,\,
\Big(\,1\,+\,{\Gamma_r\over 2L^2}\,\,x^{\alpha}\,\Gamma_{x^{\alpha}}\,\,
(1\,+\,\Gamma_{*}\,)\,\Big)\,\,\eta\,\,.
\label{adsspinor}
\eeq
The spinor $\eta$ satisfies the projections :
\beq
\Gamma_{12}\,\eta\,=\,-i\eta\,\,, \qquad\qquad
\Gamma_{34}\,\eta\,=\,i\eta\,\,,
\label{etaspinor}
\eeq
which show that this background preserves eight supersymmetries. Notice
that, since the matrix multiplying $\eta$ in eq.(\ref{adsspinor}) commutes
with $\Gamma_{12}$ and $\Gamma_{34}$, the spinor $\epsilon$ also satisfies
the conditions (\ref{etaspinor}), \ie:
\beq
\Gamma_{12}\,\epsilon\,=\,-i\epsilon\,\,, \qquad\qquad
\Gamma_{34}\,\epsilon\,=\,i\epsilon\,\,.
\label{epsilon-project}
\eeq
In eq.(\ref{adsspinor}) we are parameterizing the dependence of $\epsilon$
on the coordinates of  $AdS_5$ as in ref.\cite{LPT}. In order to explore
this dependence in detail, it is interesting to decompose the constant
spinor $\eta$ according to the different eigenvalues of the matrix
$\Gamma_*$:
\beq
\Gamma_{*}\,\eta_{\pm}\,=\,\pm\eta_{\pm}\,\,.
\label{etamasmenos}
\eeq
Using this decomposition we obtain two types of Killing spinors:
\bear
e^{{i\over 2} \psi}\,\epsilon_{-} & = & r^{1/2}\,\eta_-\,\,,\rc\rc
e^{{i\over 2} \psi}\,\epsilon_{+} & = & r^{-1/2}\,\eta_+\,+\,{r^{1/2}
\over L^2}\,\,\Gamma_r\,x^{\alpha}\, \Gamma_{x^{\alpha}}\,\eta_+\,\,.
\label{chiraladsspinor}
\eear
The four spinors $\epsilon_{-}$ are independent of the coordinates
$x^{\alpha}$ and $\Gamma_*\epsilon_{-}=-\epsilon_-$, whereas the
$\epsilon_{+}$'s do depend on the $x^{\alpha}$'s and are not eigenvectors
of $\Gamma_*$. The latter correspond to the four superconformal
supersymmetries, while the $\epsilon_{-}$'s correspond
to the ordinary ones. Notice also that the only dependence of these
spinors on the coordinates of the $Y^{p,q}$ space is through the
exponential of the angle $\psi$ in eq.(\ref{chiraladsspinor}).

In addition to the Poincare coordinates $(x^{\alpha}, r)$ used above to
represent the $AdS_5$ metric, it is also convenient to write it in the
so-called global coordinates, in which $ds^2_{AdS_5}$ takes the form:
\beq
ds^2_{AdS_5}\,=\,L^2\,\Big[ -\cosh^2\rho \,\,d\tau^2\,+\,d\rho^2\,+\,
\sinh^2\rho\,\,d\Omega_3^2 \Big]\,\,,
\label{globalADS}
\eeq
where $d\Omega_3^2$ is the metric of a unit three-sphere parameterized by
three angles $(\alpha^1, \alpha^2,\alpha^3)$:
\beq
d\Omega_3^2\,=\,(d\alpha^1)^2\,+\,\sin^2\alpha^1\Big(\,
(d\alpha^2)^2\,+\,\sin^2\alpha^2\,(d\alpha^3)^2\,\Big)\,\,,
\eeq
with $0\le\alpha^1,\alpha^2\le \pi$ and $0\le\alpha^3\le 2\pi$. In order
to write down the Killing spinors in these coordinates, we will choose
the same frame as in eq.(\ref{Ypqvielbein}) for the $Y^{p,q}$ part of the metric, while for the $AdS_5$ directions we will use:
\bear
&&e^{\tau}\,=\,L\cosh\rho\,d\tau\,\,,\,\,\,\,\,\,\,\,\,\,\,\,\,\,\,\,
e^{\rho}\,=\,Ld\rho\,\,,\rc\rc
&&e^{\alpha^1}\,=\,L\sinh\rho\,d\alpha^1\,\,,\rc\rc
&&e^{\alpha^2}\,=\,L\sinh\rho\,\sin\alpha^1\,d\alpha^2\,\,,\rc\rc
&&e^{\alpha^3}\,=\,L\sinh\rho\,\sin\alpha^1\,\sin\alpha^2\,d\alpha^3\,\,.
\eear
If we now  define the matrix
\beq
\gamma_*\,\equiv\,\Gamma_{\tau}\,\Gamma_{\rho}\,
\Gamma_{\alpha^1\,\alpha^2\,\alpha^3}\,\,,
\eeq
then, the Killing spinors in these coordinates can be written as
\cite{globalads}:
\beq
\epsilon\,=\,e^{-{i\over 2}\psi}\,
e^{-i\,{\rho\over 2}\,\Gamma_{\rho}\gamma_*}\,
e^{-i\,{\tau\over 2}\,\Gamma_{\tau}\gamma_*}\,
e^{-{\alpha^1\over 2}\,\Gamma_{\alpha^1 \rho}}\,
e^{-{\alpha^2\over 2}\,\Gamma_{\alpha^2 \alpha^1}}\,
e^{-{\alpha^3\over 2}\,\Gamma_{\alpha^3 \alpha^2}}\,\eta\,\,,
\label{globalspinor}
\eeq
where $\eta$ is a constant spinor that satisfies the same conditions as
in eq.(\ref{etaspinor}). 

\subsection{Supersymmetric probes on  $AdS_5\times Y^{p,q}$}

In the remainder of this paper we will consider Dp-brane probes moving in
the $AdS_5\times Y^{p,q}$ background. If $\xi^{\mu}$ ($\mu=0,\cdots, p$)
are a set of worldvolume coordinates and $X^M$ denote ten-dimensional coordinates, the embedding of the brane probe in the background geometry
will be characterized by the set of functions $X^M(\xi^{\mu})$, from which
the induced metric on the worldvolume is determined as: 
\beq
g_{\mu\nu}\,=\,\partial_{\mu} X^{M}\,\partial_{\nu} X^{N}\,G_{MN}\,\,,
\label{inducedmetric}
\eeq
where $G_{MN}$ is the ten-dimensional metric. Let $e^{\underline{M}}$ be
the frame one-forms of the ten-dimensional metric. These one-forms can be
written in terms of the differentials of the coordinates by means of the
coefficients $E_{N}^{\underline{M}}$:
\beq
e^{\underline{M}}\,=\,E_{N}^{\underline{M}}\,dX^N\,\,.
\eeq
From the $E_{N}^{\underline{M}}$'s and the embedding functions
$X^M(\xi^{\mu})$ we define the induced Dirac matrices on the worldvolume as:
\beq
\gamma_{\mu}\,=\,\partial_{\mu}\,X^{M}\,E_{M}^{\underline{N}}\,\,
\Gamma_{\underline{N}}\,\,,
\label{wvgamma}
\eeq
where $\Gamma_{\underline{N}}$ are constant ten-dimensional Dirac matrices.

The supersymmetric embeddings of the brane probes are obtained by imposing
the kappa-symmetry condition:
\beq
\Gamma_{\kappa}\,\epsilon\,=\,\epsilon\,\,,
\label{kappacondition}
\eeq
where $\epsilon$ is a Killing spinor of the background and $\Gamma_{\kappa}$
is a matrix that depends on the embedding. In order to write the expression
of $\Gamma_{\kappa}$ for the type IIB theory it is convenient to decompose
the complex spinor $\epsilon$ in its real and imaginary parts, $\epsilon_1$
and $\epsilon_2$. These are Majorana--Weyl spinors. They can be subsequently arranged as a two-dimensional vector
\beq
\epsilon=\epsilon_1+i\epsilon_2 ~\longleftrightarrow~ \epsilon =
\pmatrix{\epsilon_1\cr\epsilon_2}\,\,.
\eeq
The dictionary to go from complex to real spinors is straightforward,
namely:
\beq
\epsilon^* ~\longleftrightarrow~ \tau_3\,\epsilon\,\,,
\,\,\,\,\,\,\,\,\,\,\,\,\,\,\,\,\,\,\,
i\epsilon^* ~\longleftrightarrow~ \tau_1\,\epsilon\,\,,
\,\,\,\,\,\,\,\,\,\,\,\,\,\,\,\,\,\,\,
i\epsilon ~\longleftrightarrow~ -i\tau_2\,\epsilon\,\,,
\label{rule}
\eeq
where the $\tau_i$ ($i=1,2,3$) are nothing but the Pauli matrices. If there
are not worldvolume gauge fields on the D-brane, the kappa symmetry matrix
of a Dp-brane in the type IIB theory is given by \cite{swedes}:
\beq
\Gamma_{\kappa}\,=\,{1\over (p+1)!\sqrt{-g}}\,\epsilon^{\mu_1\cdots\mu_{p+1}}\,
(\tau_3)^{{p-3\over 2}}\,i\tau_2\,\otimes\,
\gamma_{\mu_1\cdots\mu_{p+1}}\,\,,
\label{gammakappa}
\eeq
where $g$ is the determinant of the induced metric $g_{\mu\nu}$ and
$\gamma_{\mu_1\cdots\mu_{p+1}}$ denotes the antisymmetrized product of 
the induced Dirac matrices (\ref{wvgamma}). To write eq.(\ref{gammakappa}) we
have assumed that the worldvolume gauge field $A$ is zero. This assumption
is consistent with the equations of motion of the probe as far as there
are no source terms in the action which could induce a non-vanishing
value of $A$. These source terms must be linear in $A$ and can only appear
in the Wess-Zumino term of the probe action, which is responsible for the
coupling of the probe to the Ramond-Ramond fields of the background. 
In the case under study only $F^{(5)}$ is non-zero (see eq.(\ref{F5}))
and the only linear term in $A$ is of the form $\int A\wedge F^{(5)}$,
which is different from zero only for a D5-brane which captures the flux
of $F^{(5)}$. This only happens for the baryon vertex configuration studied
in subsection \ref{baryonvertex}. In all other cases studied in this paper
one can consistently put the worldvolume gauge field to zero. Nevertheless,
even if one is not forced to do it, in some cases we can switch on the
field $A$ to study how this affects the supersymmetric embeddings. In
these cases the expression (\ref{gammakappa}) for $\Gamma_{\kappa}$ is
no longer valid and we must use the more general formula given in
ref.\cite{swedes}.

The kappa symmetry condition $\Gamma_{\kappa}\,\epsilon=\epsilon$ imposes
a new projection to the Killing spinor $\epsilon$ which, in general, will
not be compatible with those already satisfied by $\epsilon$ 
(see eq.(\ref{epsilon-project})). This is so because the new projections
involve matrices which do not commute with those appearing in
(\ref{epsilon-project}). The only way of making these two conditions
consistent with each other is by requiring the vanishing of the
coefficients of those non-commuting matrices, which will give rise to a
set of first-order BPS differential equations. By solving these BPS
equations we will determine the supersymmetric embeddings of the brane
probes we are looking for. Notice also that the kappa symmetry condition
must be satisfied {\em at any point of the probe worldvolume}. It is a
local condition whose global meaning, as we will see in a moment, has to
be addressed a posteriori. This requirement is not obvious at all since
the spinor $\epsilon$ depends on the coordinates (see eqs.
(\ref{adsspinor}) and (\ref{globalspinor})). However this would be
guaranteed if we could reduce the $\Gamma_{\kappa}\,\epsilon=\epsilon$
projection to some algebraic conditions on the constant spinor $\eta$ of
eqs.(\ref{adsspinor}) and (\ref{globalspinor}). The counting of solutions
of the algebraic equations satisfied by $\eta$ will give us the fraction
of supersymmetry preserved by our brane probe.

\setcounter{equation}{0}
\section{Supersymmetric D3-branes on $AdS_5\times Y^{p,q}$ }
\medskip
\label{d3}

Let us now apply the methodology just described to find the supersymmetric
configurations of a D3-brane in the $AdS_5\times Y^{p,q}$ background. 
The kappa symmetry matrix in this case can be obtained by putting $p=3$ in
the general expression (\ref{gammakappa}):
\beq
\Gamma_{\kappa}\,=\,-{i\over 4!\sqrt{-g}}\,\epsilon^{\mu_1\cdots\mu_4}\,
\gamma_{\mu_1\cdots\mu_4}\,\,,
\label{Gammad3}
\eeq
where we have used the rule (\ref{rule}) to write the expression of
$\Gamma_{\kappa}$ acting on complex spinors. 
Given that the $Y^{p,q}$ space is topologically $S^2\times S^3$, it is
natural to consider D3-branes wrapping two- and three-cycles in the
Sasaki--Einstein space. A D3-brane wrapping a two-cycle in $Y^{p,q}$
and extended
along one of the spatial directions of $AdS_5$ represents a fat string.
We will study such type of configurations in section \ref{6} where we
conclude that they are not supersymmetric, although we will find stable
non-supersymmetric embeddings of this type. 

In this section we will concentrate on the study of supersymmetric
configurations of D3-branes wrapping a three-cycle of $Y^{p,q}$.
These objects are pointlike from the gauge theory point of view and,
on the field theory side, they correspond to dibaryons constructed from
the different bifundamental fields. In what follows we  will study the
kappa symmetry condition for two different sets of worldvolume coordinates,
which will correspond to two classes of cycles and dibaryons. 

\subsection{Singlet supersymmetric three-cycles }
\label{singlet}

Let us use the global coordinates of eq.(\ref{globalADS}) to parameterize
the $AdS_5$ part of the metric and let us consider the following set of worldvolume coordinates:
\be
\xi^{\mu}=(\tau,\te,\phi,\beta),
\ee
and the following generic ansatz for the embedding:
\be
y=y(\te,\phi,\beta), \qquad \psi(\te,\phi,\beta).
\ee
The kappa symmetry matrix in this case is:
\beq
\Gamma_{\kappa}\,=\,-iL\,{\cosh\rho\over 
\sqrt{-g}}\,\Gamma_{\tau}\,\gamma_{\theta\phi\beta}\,\,.
\eeq
The induced gamma matrices along the $\theta$, $\phi$ and $\beta$ directions
can be straightforwardly obtained from (\ref{wvgamma}), namely:
\bea
{1\over L} \g_\te &=&  \frac{\sqrt{1-c\, y}}{\sqrt{6}}\,\G_3
+\frac13\psi_\te\,\G_5 -{1\over \sqrt{6}\, H}\,y_{\theta}\,\Gamma_1,
\nonumber \\
{1\over L} \g_\p &=& \frac{cH\cos\te}{\sqrt{6}}\,\G_2 + \frac{\sqrt{1-c\,
    y}}{\sqrt{6}}\sin\te\,\G_4 + \frac{1}{3} \left( \psi_\p +(1-c\,
y)\cos\te\right)\,\G_5 -{1\over \sqrt{6}\, H}\,y_{\phi}\,\Gamma_1, 
\nonumber \\
{1\over L} \g_\b &=& -\frac{H}{\sqrt{6}}\,\G_2 + \frac{1}{3}
\left(\psi_\b+y\right)\,\G_5 -{1\over \sqrt{6}\,H}\,y_{\beta}\,\G_1, 
\eea 
where the subscripts in $y$ and $\psi$ denote partial differentiation.
By using this result and the projections (\ref{epsilon-project}) the
action of the antisymmetrized product
$\gamma_{\theta\phi\beta}$ on the Killing spinor $\epsilon$ reads:
\beq
-{i\over L^3}\,\gamma_{\theta\phi\beta}\,\epsilon\,=\,
[\,a_5\,\Gamma_5\,+\,a_1\Gamma_1\,+\,a_3\Gamma_3\,+\,
a_{135}\,\Gamma_{135}\,]\,\epsilon\,\,,
\label{MSantisy}
\eeq
where the coefficients on the right-hand side are given by:
\bear
&&a_5\,=\,{1\over 18}\,\Bigg[\,(y+\psi_{\beta})\,
[\,(1-cy)\,\sin\theta\,+\,c\,y_{\theta}\cos\theta]\,+\rc\rc
&&\qquad\qquad+\,[\,\psi_\phi\,+\,(1-cy)\,\cos\theta\,]\,y_{\theta}\,-\,
\psi_\theta y_\phi\,-\,c\cos\theta 
\psi_\theta y_\beta\,\Bigg]\,\,,\rc\rc
&&a_1\,=\,-{1-cy\over 6\sqrt{6}}\,\,\sin\theta\,\big[\,
{y_\beta\over H}\,-\,iH\,\big]\,\,,\rc\rc
&&a_3\,=\,-{\sqrt{1-cy}\over  6\sqrt{6}}\,
\big[\,y_\phi\,+\,c\cos\theta y_{\beta}\,-\,i\sin\theta y_{\theta}\,
\big]\,\,,\rc\rc
&&a_{135}\,=\,{\sqrt{1-cy}\over 18}\,\,
\Bigg[\,{\sin\theta\over H}\,\big[\psi_{\theta}
y_{\beta}\,-\,(y+\psi_{\beta})\,y_\theta
\,\big]\,+\,H\big[\psi_{\phi}\,+\,(1+c\psi_\beta)\cos\theta\,\big]\,\,+\rc\rc
&&\qquad\qquad+\,\,{i\over H}\,
\Big[(\psi_{\phi}\,+\,(1-cy)\,\cos\theta\,)y_\beta\,-\,
(y+\psi_{\beta})\,y_\phi\,\Big]\,-\,iH\sin\theta\psi_{\theta}\,\,
\Bigg]\,\,.
\label{as}
\eear
As
discussed at the end of section 2, in order to implement the kappa symmetry
projection we must require the vanishing of the terms in (\ref{MSantisy})
which are not compatible with the projection (\ref{epsilon-project}). Since 
the matrices $\Gamma_1$, $\Gamma_3$ and $\Gamma_{135}$ do not commute
with those appearing in the projection (\ref{epsilon-project}), it follows
that we must impose that the corresponding coefficients vanish, \ie:
\beq
a_1 = a_3 = a_{135} = 0\,\,.
\eeq
Let us concentrate first on the condition $a_1 = 0$. By looking at its
imaginary part:
\beq
H(y)=0\,\,,
\eeq
which, in the range of allowed values of $y$, means:
\beq
y = y_1 ~, \qquad {\rm or} \qquad y = y_2 ~.
\eeq
If  $H(y)=0$, it follows by inspection
that $a_1 = a_3 = a_{135} = 0$. Notice that $\psi$ can be an arbitrary
function. Moreover, one can check that:
\beq
\left.\sqrt{-g}\right|_{BPS} \, = \, L^4\,\cosh\rho \,
\left.{a_5}\right|_{BPS}\,\,.
\eeq
Thus, one has the following equality:
\beq
\left.\Gamma_{\kappa}\,\epsilon\right|_{BPS} \, = \,
\Gamma_{\tau}\Gamma_5\,\epsilon\,\,,
\eeq
and, therefore, the condition $\Gamma_{\kappa}\epsilon=\epsilon$ becomes
equivalent to
\beq
\Gamma_{\tau}\Gamma_5\,\epsilon\,=\,\epsilon\,\,.
\label{d3-epsilon-cond}
\eeq
As it happens in the $T^{1,1}$ case \cite{acr}, the compatibility of
(\ref{d3-epsilon-cond}) with the $AdS_5$ structure of the spinor implies
that the D3-brane must be placed at $\rho=0$, \ie\ at the center of $AdS_5$.
Indeed, as discussed at the end of section \ref{ypq}, we must translate
the condition (\ref{d3-epsilon-cond}) into a condition for the constant
spinor $\eta$ of eq.(\ref{globalspinor}). Notice that $\Gamma_{\tau}\Gamma_5$
commutes with all the matrices appearing on the right-hand side of
eq.(\ref{globalspinor}) except for $\Gamma_{\rho}\gamma_*$. Since the
coefficient of $\Gamma_{\rho}\gamma_*$ in (\ref{globalspinor}) only vanishes
for $\rho=0$, it follows that only at this point the equation
$\Gamma_{\kappa}\,\epsilon=\epsilon$ can be satisfied for every point
in the worldvolume and reduces to:
\beq
\Gamma_{\tau}\Gamma_5\,\eta\,=\,\eta\,\,.
\label{d3-eta-cond}
\eeq
Therefore, if we place the D3-brane at the center of the $AdS_5$ space
 and wrap it on the three-cycles at $y=y_{1}$ or
$y=y_{2}$, we
obtain a $\frac{1}{8}$ supersymmetric configuration which preserves the
Killing spinors of the type (\ref{globalspinor}) with $\eta$ satisfying
(\ref{etaspinor}) and the additional condition (\ref{d3-eta-cond}).

The cycles we have just found have been identified by Martelli and Sparks
as those dual to the dibaryonic operators $\det (Y)$ and $\det (Z)$, made
out of the bifundamental fields that, as the D3-brane wraps the
two-sphere whose isometries are responsible for the global $SU(2)$ group,
are singlets under this  symmetry
\cite{ms}. For this reason we will refer to these cycles as singlet (S)
cycles. Let us recall how this identification is carried out. First of all,
we look at the conformal dimension $\Delta$ of the corresponding dual
operator. Following the general rule of the AdS/CFT correspondence (and
the zero-mode corrections of ref.\cite{ba2}), $\Delta=LM$, where $L$ is
given by (\ref{L}) and $M$ is the mass of the wrapped three-brane. The
latter can be computed as $M=T_3\,V_3$, with $T_3$ being the tension of
the D3-brane ($1/T_3\,=\,8\pi^3(\alpha')^2 g_s$) and $V_3$  the volume
of the three-cycle. If $g_{{\cal C}}$ is the determinant of the spatial
part of the induced metric on the three-cycle ${\cal C}$, one has:
\beq
V_3\,=\,\int_{{\cal C}}\sqrt{g_{{\cal C}}}\,\,\,d^3\xi\,\,.
\eeq
For the singlet cycles ${\rm S}_i$ at $y=y_i$ ($i=1,2$) and $\psi$=constant,
the volume $V_3$ is readily computed, namely:
\beq
V_3^{{\rm S}_i}\,=\,{2L^3\over 3}\,(\,1-cy_i\,)\,|\,y_i\,|\,
(2\pi)^2\,\ell\,\,.
\eeq
Let us define $\lambda_1=+1$, $\lambda_2=-1$. Then, if
$\Delta_i^{{\rm S}}\,\equiv\,\Delta^{{\rm S}_i}$, one has:
\beq
\Delta_i^{{\rm S}}\,=\,{N\over 2q^2}\,\Big[\,-4p^2+3q^2+2\lambda_i\,pq\,+\,
(2p-\lambda_i\,q)\,\sqrt{4p^2\,-\,3q^2}\,\,\Big]\,\,.
\eeq
As it should be for a BPS saturated object, the R-charges $R_i$ of the
${\rm S}_i$ cycles are related to $\Delta_i^{{\rm S}}$ as $R_i={2\over 3}\,
\Delta_i^{{\rm S}}$. By comparing the values of $R_i$ with those determined
in \cite{sequiver} from the gauge theory dual (see Table \ref{charges})
one concludes that, indeed, a D3-brane wrapped at $y=y_1$ ($y=y_2$) can be
identified with the operator $\det (Y)$ ($\det (Z)$) as claimed. Another
piece of evidence which supports this claim is the calculation of the
baryon number, that can be identified with the third homology class of
the three-cycle ${\cal C}$ over which the D3-brane is wrapped. This number
(in units of $N$) can be obtained by computing the integral over ${\cal C}$
of the pullback of a $(2,1)$ three-form $\Omega_{2,1}$ on $CY^{p,q}$:
\beq
{\cal B}({\cal C})\,=\,\pm i \int_{{\cal C}}\,
P\big[\,\Omega_{2,1}\,\big]_{{\cal C}}\,\,,
\label{baryon-def}
\eeq
where $P[\cdots]_{{\cal C}}$ denotes the pullback to the cycle ${{\cal C}}$
of the form that is inside the brackets. The sign of the right-hand side
of (\ref{baryon-def}) depends on the orientation of the cycle. The explicit
expression of $\Omega_{2,1}$ has been determined in ref.\cite{hek}:
\beq
\Omega_{2,1}\,=\,K\,\Big(\,{dr\over r}\,+\,\frac{i}{L}\,
e^5\,\Big)\wedge \omega\,\,,
\eeq
where $e^5$ is the one-form of our vielbein (\ref{Ypqvielbein}) for the
$Y^{p,q}$ space, $K$ is the constant
\beq
K\,=\,{9\over 8\pi^2}\,(p^2-q^2)\,\,,
\eeq
and $\omega$ is the  two-form:
\beq
\omega\,=\,-{1\over (1-cy)^2\, L^2}\,\,\Big[\,e^1\wedge e^2\,+\,e^3\wedge
e^4\,\Big]\,\,.
\eeq
Using $(\theta,\phi,\beta)$ as worldvolume coordinates of the singlet
cycles ${\rm S}_i$,
\beq
P\big[\,\Omega_{2,1}\,\big]_{{\rm S}_i}\,=\,-i\,{K\over 18}\,
{y_i\over 1-cy_i}\,\sin\theta\, d\theta\wedge d\phi\wedge d\beta\,\,,
\eeq
Then,
changing variables from $\beta$ to $\alpha$ by means of
(\ref{beta-alpha}), and taking into account that $\alpha\in [0,2\pi
\ell]$, one gets:
\beq
\int_{{\rm S}_i}\, P\big[\,\Omega_{2,1}\,\big]_{{\rm S}_i}\,=\,-i\,
{8\pi^2\over 3}\, {K \ell y_i\over 1 - c y_i}\,\,.
\eeq
After using the values of $y_1$ and $y_2$ displayed in (\ref{y12}), we
arrive at:
\bear
&&{\cal B}({\rm S}_1)\,=\,- i \int_{{\rm S}_1}\,
P\big[\,\Omega_{2,1}\,\big]_{{\rm S}_1}\,=\,p-q\,\,,\rc\rc
&&{\cal B}({\rm S}_2)\,=\, i \int_{{\rm S}_2}\,
P\big[\,\Omega_{2,1}\,\big]_{{\rm S}_2}\,=\,p+q\,\,.
\eear
Notice the perfect agreement of ${\cal B}({\rm S}_1)$ and ${\cal B}({\rm
S}_2)$ with the baryon numbers of $Y$ and $Z$ displayed in Table
\ref{charges}. 

\subsection{Doublet  supersymmetric three-cycles }
\label{hekconst}

Let us now try to find supersymmetric embeddings of D3-branes on three-cycles
by using a different set of worldvolume coordinates. As in the previous subsection it is convenient to use the global coordinates (\ref{globalADS})
for the $AdS_5$ part of the metric and the following set of worldvolume
coordinates:
\be
\xi^\mu=(\tau,y,\beta,\psi) ~.
\ee
Moreover, we will adopt the ansatz:
\be
\theta(y,\b, \psi) ~, \qquad \phi(y,\b,\psi) ~.
\label{embdoublet}
\ee
The kappa symmetry matrix $\G_{\k}$ in this case takes the form:
\be
\G_{\k}=-iL\,{\cosh\rho\over 
\sqrt{-g}}\,\Gamma_{\tau}\,\gamma_{y\,\beta\,\psi}\,\,,
\ee
and the induced gamma matrices are:
\bea
&&{1\over L}\g_y = -\frac{1}{\sqrt{6}H} \G_1 +
\frac{cH\cos\te}{\sqrt{6}} \p_y \G_2 + \frac{\sqrt{1-c
y}}{\sqrt{6}} \left( \te_y \G_3 +
\p_y\sin\theta\, \G_4 \right) + \frac{1-c\, y}{3}\cos\te \p_y\, \G_5\,\,,
\nonumber \\ \nonumber \\
&&{1\over L}\g_\b= \frac{H}{\sqrt{6}}\left(-1+ c\cos\te\,
\p_\b\right)\, \G_2  +  \frac{\sqrt{1-c\, y}}{\sqrt{6}}\,\te_\b\, \G_3 
+  \frac{\sqrt{1-c\, y}}{\sqrt{6}}\sin\te \,\p_\b\, \G_4 \nonumber\\
\nonumber\\
&&\qquad\qquad+ {1\over 3}\Big( \,y +
(1-c\, y) \cos\te\, \p_\b \Big)\, \G_5\,\,,  \\ \nonumber \\
&&{1\over L}\g_\psi= \frac{cH\cos\te}{\sqrt{6}}\,\p_\psi\, \G_2 +
\frac{\sqrt{1-c\, y}}{\sqrt{6}}\, \left( \te_\psi\, \G_3 + \sin\te\,
\p_\psi \,\G_4 \right) + {1\over 3}\left(1 +(1-c)\,
\cos\te\p_\psi\right)\, \G_5\,\,.
\nonumber
\eea
By using again the projections (\ref{epsilon-project}) one easily gets
the action of $\gamma_{y\,\beta\,\psi}$ on the Killing spinor
\beq
-{i\over L^3}\,\gamma_{y\,\beta\,\psi}\,\epsilon\,=\,
\big[\,c_5\,\Gamma_5\,+\,c_1\,\Gamma_1\,+\,c_3\,\Gamma_3\,+\,
c_{135}\,\Gamma_{135}\,\big]\,\epsilon\,\,,
\label{HEKcs}
\eeq
where the different coefficients appearing on the right-hand side of 
(\ref{HEKcs}) are given by:
\bear
&&c_5 = {1\over 18}\,\Bigg[ -1 - \cos\theta (\phi_\psi - c \phi_{\beta}) +
(1-cy) \sin\theta\, \Big[ \theta_y (\phi_\beta - y\phi_\psi) -
\phi_y (\theta_\beta - y\theta_\psi)\,\Big]\,\Bigg]\,,\rc\rc
&&c_1\,=\,-{1-cy\over 6\sqrt{6}}\,\sin\theta\,
\Big[\,{\theta_{\beta}\phi_{\psi}-\theta_{\psi}\phi_{\beta}\over H}\,+\,
iH\,(\theta_y\phi_{\psi}\,-\,\theta_\psi\phi_{y})\,\Big]\,\,,\rc\rc
&&c_3\,=\,-{\sqrt{1-cy}\over 6\sqrt{6}}\,\Big[\,
\theta_{\psi}\,-\,c\cos\theta\,(\theta_{\psi}\phi_\beta\,-\,
\theta_\beta\phi_\psi)\,+\,i\sin\theta\phi_{\psi}\,\Big]\,\,,\rc\rc
&&c_{135} = -{\sqrt{1-cy}\over 18}\,\Bigg[ {\sin\theta\over H}
(\phi_\beta - y\phi_\psi) + H \bigg( \theta_y + \cos\theta \bigg[
\theta_y (\phi_{\psi} - c\phi_\beta) - \phi_y (\theta_{\psi} -
c\theta_\beta) \bigg] \bigg) \rc\rc
&&\qquad\qquad + iH\sin\theta\,\phi_y -
{i\over H}\,\Big[ \theta_{\beta} - y \theta_\psi + (1-cy) \cos\theta
(\theta_\beta \phi_\psi - \theta_{\psi} \phi_\beta) \Big] \Bigg]\,.
\label{expressionHEKcs}
\eear
Again, we notice that the matrices $\Gamma_1$, $\Gamma_3$ and $\Gamma_{135}$
do not commute with the projections (\ref{epsilon-project}). We must impose:
\beq
c_1=c_3=c_{135}=0\,\,.
\eeq
From the vanishing of the imaginary part of $c_3$ we obtain the condition:
\beq
\sin\theta\,\phi_\psi\,=\,0\,\,.
\label{HEKimc3}
\eeq
One can solve the condition (\ref{HEKimc3}) by taking $\sin\theta\,=\,0$,
\ie\ for $\theta=0,\pi$. By inspection one easily realizes that $c_1$, $c_3$
and $c_{135}$ also vanish for these values of $\theta$ and for an arbitrary
function $\phi(y,\beta,\psi)$. Therefore, we have the solution
\beq
\theta=0,\pi\,\,,\qquad \phi=\phi(y,\beta,\psi)\,\,.
\eeq
Another possibility is to take $\phi_\psi=0$. In this case one readily
verifies that $c_1$ and $c_3$ vanish if $\theta_\psi=0$. Thus, let us
assume that both $\phi$ and $\theta$ are independent of the angle $\psi$.
From the vanishing of the real and imaginary parts of $c_{135}$ we get two
equations for the functions $\theta=\theta(y,\beta)$ and
$\phi=\phi(y,\beta)$, namely:
\bear
&&\theta_y\,+\,
{\sin\theta\over H^2}\,\phi_{\beta}\,+\,c\cos\theta\,
(\phi_y\,\theta_\beta\,-\,\theta_y\,\phi_\beta)\,=\,0\,\,,\rc\rc
&&\theta_\beta\,-\,H^2\,
\sin\theta\phi_y\,=\,0\,\,.
\label{HEKBPS}
\eear
If the BPS equations (\ref{HEKBPS}) hold, one can verify that the kappa
symmetry condition $\Gamma_{\kappa}\,\epsilon\,=\,\epsilon$ reduces, up
to a sign, to the projection (\ref{d3-epsilon-cond}) for the Killing
spinor. As in the case of the S three-cycles studied in subsection
\ref{singlet}, by using the explicit expression (\ref{globalspinor}) of
$\epsilon$ in terms of the global coordinates of $AdS_5$, one concludes
that the D3-brane must be placed at $\rho=0$. The corresponding
configuration preserves four supersymmetries. 

In the next subsection we will tackle the problem of finding the general
solution of the system (\ref{HEKBPS}). Here we will analyze the trivial
solution of this system, namely:
\beq
\theta\,=\,{\rm constant}\,\,,\qquad\qquad
\phi\,=\,{\rm constant}\,\,.
\label{HEKconstant}
\eeq
This kind of three-cycle was studied in ref.\cite{hek} by Herzog, Ejaz and Klebanov (see also \cite{sequiver}), who showed that it corresponds to
dibaryons made out of the $SU(2)$
doublet fields $U^{\alpha}$. In what follows we will refer to it as doublet
(D) cycle. Let us review the arguments leading to this identification. First
of all, the volume of the D cycle (\ref{HEKconstant}) can be computed
with the result:
\beq
V_3^{D}\,=\,{L^3\over 3}\,(2\pi)^2\, (y_2-y_1)\,\ell\,\,.
\eeq
By using the values of $y_{1}$ and $y_{2}$ (eq.(\ref{y12})), $L$
(eq.(\ref{L})) and 
$\ell$ (eq.(\ref{alphaperiod})) we find the following value of the conformal
dimension:
\beq
\Delta^{D}\,=\,N\,{p\over q^2}\,\Big(\,2p\,-\,\sqrt{4p^2 - 3q^2}\,\,\Big)\,.
\eeq
By comparison with Table \ref{charges}, one can verify that the
corresponding R-charge, namely $2/3\, \Delta^{D}$, is equal to the R-charge of
the field $U^{\alpha}$ multiplied by $N$. We can check this identification
by computing the baryon number. Since, in this case, the pullback of
$\Omega_{2,1}$ is:
\beq
P\big[\,\Omega_{2,1}\,\big]_{{\rm D}}\,=\,i\,{K\over 3
(1-cy)^2}\,dy\wedge\,d\alpha\wedge d\psi
\eeq
we get:
\beq
{\cal B}({\rm D})\,=\,- i \int_{{\rm D}}\,
P\big[\,\Omega_{2,1}\,\big]_{{\rm D}}\,=\,-p\,\,,
\eeq
which, indeed, coincides with the baryon number of $U^{\alpha}$ written in
Table \ref{charges}. 

\subsubsection{General integration}

Let us now try to integrate in general the first-order differential system
(\ref{HEKBPS}). With this purpose it is more convenient to describe the
locus of the D3-brane by means of two functions $y = y(\theta,\phi)$,
$\beta =
\beta(\theta,\phi)$. Notice that this is equivalent to the description used
so far (in which the independent variables were $(y, \beta)$), except for
the cases in which $(\theta,\phi)$ or $(y, \beta)$ are constant. The
derivatives in these two descriptions are related by simply inverting the
Jacobian matrix, \ie:
\beq
\pmatrix{ y_\theta & y_\phi \cr \beta_\theta &
\beta_\phi}\,=\,\pmatrix{ \theta_y &\theta_\beta \cr \phi_y &
\phi_\beta}^{-1}\,\,.
\eeq
By using these equations the first-order system (\ref{HEKBPS}) is equivalent
to:
\beq
\beta_\theta\,=\,{y_\phi\over H^2\sin\theta}\,\,,
\qquad\qquad
\beta_\phi\,=\,c\cos\theta\,-\,{\sin\theta\over H^2}\,y_{\theta}\,\,.
\label{BPSthetaphi}
\eeq
These equations can be obtained directly by using $\theta$ and $\phi$ as
worldvolume coordinates. Interestingly, in this form the BPS equations can
be written as Cauchy--Riemann equations and, thus, they can be integrated in
general. This is in agreement with the naive expectation that, at least
locally, these equations should determine some kind of holomorphic
embeddings. In order to verify this fact, let us introduce new
variables $u_1$ and $u_2$, related to $\theta$ and $y$ as follows:
\begin{equation}
 u_1=\,\log\,
\bigg(\,\tan
\frac{\theta}{2}\,\bigg), \qquad \qquad u_2=\,\log\, \bigg(\,\frac{(\sin
\theta)^c}{f_1(y)}\,\bigg).
\end{equation}
By comparing the above expressions with the definitions of $z_1$ and $z_2$
in eq.(\ref{complexzs}), one gets:
\beq
u_1-i\phi\,=\,\log z_1\,\,, \qquad\qquad
u_2-i\beta\,=\,\log z_2\,\,.
\eeq
The relation between $u_1$ and $\theta$ leads to $du_1 =
d\theta/\sin\theta$, from which it follows that:
\beq
{\partial u_2\over \partial u_1}\,=\,c\cos\theta - {\sin\theta\over
H^2}\,y_{\theta}\,\,, \qquad\qquad\qquad
{\partial \beta\over \partial u_1}\,=\,\sin\theta\,\beta_{\theta}\,\,,
\eeq
and it is easy to demonstrate that the BPS equations (\ref{BPSthetaphi})
can be written as:
\begin{equation}
\frac{\partial u_2}{\partial u_1} = \frac{\partial \beta}{\partial
\phi}\,, \qquad\qquad \frac{\partial u_2}{\partial \phi} = -
\frac{\partial \beta}{\partial u_1}\,,
\end{equation}
these being the Cauchy--Riemann equations for the variables $u_2 - i\beta
= \log z_2$ and $u_1 - i\phi = \log z_1$. Then, the general integral of the
BPS equations is
\begin{equation}
\log z_2\,=\,f(\log z_1)\,\,,
\label{logrelation}
\end{equation}
where $f$ is an arbitrary (holomorphic) function of $\log z_1$. By
exponentiating eq.(\ref{logrelation}) one gets that the general solution of
the BPS equations is a function $z_2=g(z_1)$, in which $z_2$ is an arbitrary
holomorphic function of $z_1$. This result is analogous to what happened for
$T^{1,1}$ \cite{acr}. The appearance of a holomorphic function in the local
complex coordinates $z_1$ and $z_2$ is a consequence of kappa symmetry or,
in other words, supersymmetry. But one still has to check that this equation
makes sense globally. We will come to this point shortly. The simplest case
is that in which $\log z_2$ depends linearly on $\log z_1$, namely
\begin{equation}
\log z_2\,=\,n(\log z_1)\,+\,{\rm const.} ~,
\end{equation}
where $n$ is a constant. By exponentiating this equation we get a relation between 
$z_2$ and $z_1$ of the type:
\beq
z_2\,=\,{\cal C}\,z_1^n ~,
\label{n-winding}
\eeq
where ${\cal C}$ is a complex constant. If we represent this constant as
${\cal C}=Ce^{-i\beta_0}$, the embedding (\ref{n-winding}) reduces to the
following real functions $\beta=\beta(\phi)$ and $y=y(\theta)$:
\bear
\beta&=&n\phi\,+\,\beta_0\,\,,\rc\rc
f_1(y)&=&C\,\,{\big (\,\sin\theta\,\big)^c\over \Big(\tan{\theta\over
2}\Big)^n}\,\,.
\label{embgen}
\eear
This is a nontrivial embedding of a probe D3-brane on $AdS_5 \times Y^{p,q}$.
Notice that in the limit $c \to 0$ one recovers the results of \cite{acr}.
For $c \neq 0$, a key difference arises. As we discussed earlier, $z_2$ is
not globally well-defined in $CY^{p,q}$ due to its dependence on $\beta$. 
As a consequence, eqs.(\ref{n-winding})--(\ref{embgen}) describe a
kappa-symmetric embedding for the D3-brane on $Y^{p,q}$ but it does not
correspond to a wrapped brane. The D3-brane spans a submanifold with
boundaries.\footnote{In this respect, notice that it might happen that
global consistency forces, through boundary conditions, the D3-brane
probes to end on other branes.} The only solution corresponding to a probe
D3-brane wrapping a three-cycle is $z_1 = {\rm const.}$ which is the one
obtained in the preceding subsection.

In order to remove $\beta$ while respecting holomorphicity \footnote{One
might think that a possible caveat to this problem is to choose a different
slicing of $Y^{p,q}$ as the one in (\ref{alphametric}), where the metric is
written as a U(1) bundle coordinatized by $\alpha$ (the base not being a
K\"ahler--Einstein manifold). The complex coordinates of the slice are
\beq
\tilde z_1 = z_1 ~, ~~~~~~~ \tilde z_2 = G(y) \sin\te e^{i \psi} ~,
\eeq
where $G'(y)/G(y) = 3/\sqrt{w(y)} q(y)$. However, a `holomorphic' ansatz of
the form $\tilde z_1 = \tilde z_2^m$ would be related to an embedding of
the form
$\phi = \phi(\psi)$ and $\te = \te(y)$, which is a particular case of
(\ref{embdoublet}) albeit it is not kappa symmetric. These complex
coordinates $\tilde z_1$ and $\tilde z_2$ have nothing to do with the
complex structure of the Calabi-Yau manifold and, as such, kappa symmetry
is not going to lead to a holomorphic embedding in terms of them.}, we
seem to be forced to let $z_3$ enter into the game. The reason is simple,
any dependence in $\beta$ disappears if $z_2$  enters through the
product $z_2 z_3$. This would demand embeddings involving the radius that
we did not consider. In this respect, it is interesting to point out that
this is also the conclusion reached in \cite{BHOP} from a different perspective:
there, the complex coordinates corresponding to the generators of the
chiral ring are deduced and it turns out that all of them depend on $z_1$,
$z_2 z_3$ and $z_3$. It would be clearly desirable to understand these
generalized wrapped D3-branes in terms of algebraic geometry, following
the framework of Ref.\cite{beasley} which, in the case of the conifold, 
emphasizes the use of global homogeneous coordinates. Unfortunately, the
relation between such homogeneous coordinates and the chiral fields of
the quiver theory is more complicated in the case of $CY^{p,q}$.

\subsection{The calibrating condition}

Let us now verify that the BPS equations we have obtained ensure that the
three-dimensional submanifolds we have found are calibrated. First, recall
that the metric of the $Y^{p,q}$ manifold can be written as (\ref{slice}),
$$ ds^2_{Y^{p,q}}\,=\,ds^2_{4} + \left[ \frac{1}{3} d\psi + \sigma
\right]^2 ~, $$
where $\sigma$ is a one-form given by
\beq
\sigma\,=\,{1\over 3}\,\big[\,\cos\theta d\phi\,+\,y (d\beta\,-\,
c\cos\theta d\phi)\,\big]\,\,.
\eeq
The  K\"ahler form $J_4$ of the four-dimensional K\"ahler-Einstein space
is just
\beq
J_4\,=\,{1\over 2}\,d\sigma\,=\,{1\over L^2}\,\big[\,
e^1\wedge e^2\,-\,e^3\wedge e^4\,\big]\,\,,
\eeq
where the $e^i$'s are the forms of the vielbein (\ref{Ypqvielbein}). From
the Sasaki-Einstein space $Y^{p,q}$  we can construct the Calabi-Yau cone
$CY^{p,q}$, whose metric is just given by: $ds^2_{CY^{p,q}}\,=\,dr^2 +
r^2\,ds^2_{Y^{p,q}}$. The  K\"ahler form $J$ of $CY^{p,q}$ is just:
\beq
J\,=\,r^2\,J_4\,+\,{r\over L}\,dr\wedge e^5\,\,,
\eeq
whose explicit expression in terms of the coordinates is:
\beq
J\,=\,-{r^2\over 6}\,(1-cy)\,\sin\theta d\theta\wedge d\phi\,+\,{1\over
3}\,r dr\wedge (d\psi+\cos\theta d\phi)\,+\,{1\over 6} d(r^2y)\wedge 
(d\beta\,-\,c\cos\theta d\phi)\,\,.
\eeq
Given a three-submanifold in $Y^{p,q}$ one can construct its cone
${\cal D}$, which is a four-dimensional submanifold of $CY^{p,q}$. The
calibrating condition for a supersymmetric four-submanifold ${\cal D}$ of
$CY^{p,q}$ is just:
\beq
P\Big[\,{1\over 2}\,J\wedge J\,\Big]_{{\cal D}}\,=\,
{\rm Vol}({\cal D})\,\,,
\label{calibration}
\eeq
where  ${\rm Vol}({\cal D})$ is the volume form of the divisor 
${\cal D}$. Let us check that the condition (\ref{calibration}) is indeed
satisfied by the cones constructed from our three-submanifolds. In order
to verify this fact it is more convenient to describe the embedding by
means of functions
$y=y(\theta,\phi)$ and $\beta=\beta(\theta,\phi)$. The corresponding BPS
equations are the ones written in (\ref{BPSthetaphi}). By using them one
can verify that the induced volume form for the three-dimensional
submanifold is:
\beq
vol\,=\,{1\over 18}\,\Big|\,(1-cy)\sin\theta\,+\,c\cos\theta y_{\theta}\,+\,
\beta_\theta y_\phi\,-\,y_\theta\beta_\phi\,\Big|_{BPS}\,\,
d\theta\wedge d\phi\wedge d\psi ~.
\eeq
By computing the pullback of $J\wedge J$ one can verify that the calibrating condition
(\ref{calibration}) is indeed satisfied for:
\beq
{\rm Vol}({\cal D})\,=\,-r^3\,dr\wedge vol\,\,,
\eeq
which is just the volume form of ${\cal D}$ with the metric
$ds^2_{CY^{p,q}}$ having a particular orientation. Eq. (\ref{calibration})
is also satisfied for the cones constructed from the singlet and doublet 
three-cycles of sections \ref{singlet} and \ref{hekconst}. This fact is
nothing but the expression of the local nature of supersymmetry.

\subsection{Energy bound}

The dynamics of the D3-brane probe is governed by the Dirac-Born-Infeld lagrangian that, for the case in which there are not worldvolume gauge
fields, reduces to:
\beq
{\cal L}\,=\,-\sqrt{-g}\,\,,
\label{D3lag-den}
\eeq
where we have taken the D3-brane tension equal to one. We have checked that any solution of the first-order
equations (\ref{HEKBPS}) or (\ref{BPSthetaphi}) also satisfies the
Euler-Lagrange equations derived from the lagrangian density
(\ref{D3lag-den}). Moreover, for the static configurations we are
considering here the hamiltonian density ${\cal H}$  is, as expected, 
just ${\cal H}=-{\cal L}$. We are now going to verify that this energy
density satisfies a bound, which is just saturated when the BPS equations
(\ref{HEKBPS}) or (\ref{BPSthetaphi}) hold. In what follows we will take
$\theta$ and $\phi$ as independent variables. For an  arbitrary 
embedding of a D3-brane described by two functions $\beta=\beta(\theta,
\phi)$ and
$y=y(\theta, \phi)$ one can show that ${\cal H}$ can be written as:
\beq
{\cal H}\,=\,\sqrt{{\cal Z}^2\,+\, {\cal Y}^2\,+{\cal W}^2}\,\,,
\eeq
where ${\cal Z}$, ${\cal Y}$ and ${\cal W}$ are given by:
\bear
{\cal Z}&=&{L^4\over 18}\,\Bigg[\,
(1-cy)\sin\theta\,+c\cos\theta y_{\theta}+ 
y_{\phi}\beta_{\theta}-y_{\theta}\beta_{\phi}\,\Bigg]\,\,,\rc\rc
{\cal Y}&=&{L^4\over 18}\,\sqrt{1-cy}\,\,H\,
\Bigg[\,\beta_{\phi}\,-\,c\cos\theta\,+\,{\sin\theta \over H^2}\,y_{\theta}
\,\Bigg]\,\,,\rc\rc
{\cal W}&=&{L^4\over 18}\,\sqrt{1-cy}\,\,H\,
\Bigg[\,\sin\theta\,\beta_{\theta}\,-\,{y_{\phi}\over H^2}\,\Bigg]\,\,.
\eear
Obviously one has:
\beq
{\cal H}\ge \big|\,{\cal Z}\,\big|\,\,.
\label{energybound}
\eeq
Moreover, since 
\beq
{\cal Y}_{\big |\,BPS}\,=\,{\cal W}_{\big |\,BPS}\,=\,0\,\,,
\eeq
the bound saturates when the BPS equations (\ref{BPSthetaphi}) are satisfied.
Thus, the system of
differential equations (\ref{BPSthetaphi}) is equivalent to the condition 
${\cal H}= \big|\,{\cal Z}\,\big|$ (actually ${\cal Z}\ge 0$ if the BPS
equations (\ref{BPSthetaphi}) are satisfied). Moreover, for an arbitrary
embedding
${\cal Z}$ can be written as a total derivative, namely:
\beq
{\cal Z}\,=\,{\partial\over \partial \theta}\,{\cal Z}^{\theta}\,+\,
{\partial\over \partial \phi}\,{\cal Z}^{\phi}\,\,.
\eeq
This result implies that ${\cal H}$ is bounded by the integrand of a topological
charge. The explicit form of ${\cal Z}^{\theta}$ and ${\cal Z}^{\phi}$ is:
\bear
{\cal Z}^{\theta}&=&-{L^4\over 18}\,\Big[\,(1-cy)\cos\theta\,+\,y\,\beta_{\phi}\,
\Big]\,\,,\rc\rc
{\cal Z}^{\phi}&=&{L^4\over 18}\,y\,\beta_{\theta}\,\,.
\eear
In this way, from the point of view of the D3-branes, the configurations
satisfying eq. (\ref{BPSthetaphi})  can be regarded as BPS worldvolume
solitons.

\subsection{BPS fluctuations of dibaryons}

In this section we study BPS fluctuations of dibaryon
operators in the $Y^{p,q}$ quiver theory. We start with the
simplest dibaryon which is singlet under $SU(2)$, say  $\det Y$. 
To construct excited dibaryons we should replace one of the $Y$ factors 
by any other chiral field  transforming in the same
representation of the gauge groups. For example,  replacing $Y$
by
$YU^{\alpha}V^{\beta}Y$, we get a new operator of the form
\begin{eqnarray}
\epsilon_1\epsilon^2(YU^{\alpha}V^{\beta}Y)Y\cdots Y\,\,,
\end{eqnarray}
where $\epsilon_1$ and $\epsilon^2$ are abbreviations for the
completely anti-symmetric tensors for the respective $SU(N)$
factors of the gauge group. Using the identity
\begin{eqnarray}
\epsilon^{a_1\cdots a_N}\epsilon_{b_1\cdots
b_N}=\sum_{\sigma}(-1)^{\sigma}\delta^{a_1}_{\sigma(b_1)}\cdots
\delta^{a_N}_{\sigma(b_N)}\,\,,
\end{eqnarray}
the new operator we get can factorize into the original dibaryon
and a single-trace operator
\begin{eqnarray}
{\rm Tr}(U^{\alpha}V^{\beta}Y)\;{\rm det}\,Y\,\,.
\end{eqnarray}
Indeed for singlet dibaryons, a factorization of this sort always works.
This fact seems to imply, at least at weak coupling,  that excitation of a
singlet dibaryon can be represented as graviton fluctuations in the
presence of the original dibaryon.

For the case of dibaryon with $SU(2)$ quantum number the
situation is different. Consider, for simplicity, the state with
maximum $J_3$ of the $SU(2)$
\begin{eqnarray}
\epsilon_1\epsilon^2(U^1\cdots U^1)={\rm det}\,U^1,
\end{eqnarray}
we can replace one of $U^1$ factors by $U^1\, {\cal O}$, where $\cal O$ is
some operator given by a closed loop in the quiver. As the case of
singlet dibaryon, this kind of excitation is factorizable since
all $SU(2)$ indices are symmetric. So this kind of operator should
be identified with a graviton excitation with wrapped D3-brane in the
dual string theory. However if the $SU(2)$ index of the $U$ field
is changed in the excitation, i.e. $U^1\rightarrow U^2\, {\cal O}$,
then the resulting operator cannot be written as a product of the
original dibaryon and a meson-like operator. Instead it
has to be interpreted as a single particle state in $AdS$. Since
the operator also carries the same baryon number, the natural
conclusion is that the one-particle state is a BPS excitation of
the wrapped D3-brane corresponding to the dibaryon \cite{ba2}.

\begin{figure}[ht]
\centerline{\epsffile{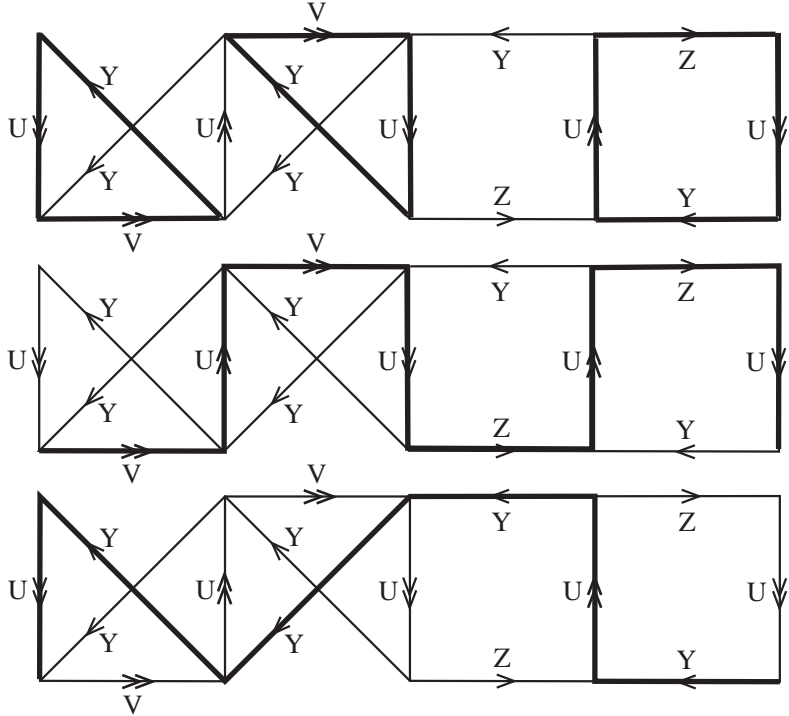}}
\caption{Loops in the $Y^{4,2}$ quiver representing mesonic operators
in the chiral ring. There are short loops such as $UVY$, $VUY$ or $YUZU$
(upper), longest loops as $VUVUZUZU$ (middle) and long loops like $YUYYYU$
(bottom). The representative of each class in the chiral ring is,
respectively, ${\cal O}_1$, ${\cal O}_2$ and ${\cal O}_3$.}
\label{Y42loops}
\end{figure}

In order to classify all these BPS excitations of the dibaryon, we
have to count all possible inequivalent chiral operators $\cal O$
that transform in the bifundamental representation of one of the
gauge group factors of the theory. In $Y^{p,q}$ quiver gauge theory, these
operators correspond to loops in the quiver diagram just like the mesonic
chiral operators discussed in \cite{bk}. The simplest ones are
operators with R-charge 2. They have been thoroughly discussed in
\cite{benami}. They are given by short loops of length 3 or 4 in
the quiver, precisely as those operators entering in the superpotential
(\ref{supYpq}). They are single trace operators of the form
(in what follows we omit the trace and the $SU(2)$ indices) $UVY$,
$VUY$ or $YUZU$ (see the upper quiver in Fig.\ref{Y42loops}).
Since they are equivalent in the chiral ring, we can identify
them as a single operator ${\cal O}_1$. It transforms in the
spin $\frac{1}{2} \otimes \frac{1}{2} = 0 \oplus 1$ representation
of the global $SU(2)$. The scalar component vanishes in the chiral
ring. Thus, we end up with a spin 1 chiral operator with scaling
dimension $\Delta = 3$. Its $U(1)_F$ charge vanishes.

There are also two classes of long loops in the quiver. The first
class, whose representative is named ${\cal O}_2$, has length $2p$,
winds the quiver from the left to the right and is made of $p$ $U$
type operators, $q$ $V$ type operators and $p-q$ $Z$ type operators.
For example, in $Y^{4,2}$, a long loop of this class is $VUVUZUZU$
(middle quiver in Fig.\ref{Y42loops}). It transforms in the spin
$\frac{1}{2} \otimes ... \otimes \frac{1}{2} = \frac{p+q}{2} \oplus
\dots$ representation of $SU(2)$. The dots amount to lower dimensional
representations that vanish in the chiral ring. The resulting operator,
${\cal O}_2$, has spin $\frac{p+q}{2}$. There is another class of
long loops which has length $2p-q$, running along the quiver in the
opposite direction, build with $p$ $Y$ type operators and $p-q$ $U$
type operators. We name its representative as ${\cal O}_3$. 
In the case of $Y^{4,2}$, it is an operator like $YUYYYU$ (bottom
quiver in Fig.\ref{Y42loops}). $SU(2)$ indices, again, have to be
completely symmetrized, the spin being $\frac{p-q}{2}$. Long loops
wind around the quiver and this leads to a nonvanishing value of
$Q_F$ \cite{bk}. The baryonic charge vanishes for any of these loops.
We summarize in Table \ref{chring} the charge assignments for the
three kinds of operators ${\cal O}_i$ \cite{bk}.
\begin{table}
\begin{center}
$$\begin{array}{|c|c|c|c|}
\hline {\rm Operator} & Q_R & Q_F & {\rm Spin} \\
\hline\hline
& & & \\[-1.5ex]
{\cal O}_1 & 2 & 0 & 1 \\[1ex]
\hline
& & & \\[-1.5ex]
{\cal O}_2 & p+q-\frac{1}{3\ell} & p & \frac{p+q}{2}\\[1ex]
\hline
& & & \\[-1.5ex]
{\cal O}_3 & p-q+\frac{1}{3\ell} & -p & \frac{p-q}{2}\\[1ex]
\hline
\end{array}$$
\caption{Charges assignments for the mesonic operators
${\cal O}_i$ that generate the chiral ring.}
\label{chring}
\end{center}
\end{table}
We can see that these operators satisfy the BPS condition $\Delta={3\over
2}\, Q_R$. In fact, they are the building blocks of all other scalar BPS
operators. The general BPS excitation corresponds to operators of the form
\begin{eqnarray}
{\cal O}=\prod_{i=1}^3{\cal O}_{i}^{\,\,n_i}\,\,.
\end{eqnarray}
It is interesting to notice that the spectrum of  fluctuations
of a dibaryon must coincide with the mesonic chiral operators in the 
$Y^{p,q}$ quiver theory. This would provide a nontrivial test of the 
AdS/CFT correspondence. We show this result explicitly via an analysis of
open string fluctuation on wrapped  D3-branes.

Now we are interested in describing the excitations of dibaryon operators from the dual 
string theory. For those excitations that are
factorizable, the dual configurations are just the multi-particle
states of graviton excitations in the presence of a dibaryon. The
correspondence of graviton excitation and mesonic operator were
studied in \cite{bk}\cite{ksy}. What we are really interested in
are those non-factorizable operators that can be interpreted as
open string excitations on the D-brane. This can be analyzed by
using the Dirac-Born-Infeld action of the D3-brane. In what follows  we will
focus on the dibaryon made of $U$ fields, which corresponds to the 
three-cycle D studied in subsection \ref{hekconst} which, for convenience,
we will  parameterize with the coordinates $(y,\psi, \alpha)$. The
analysis of the dibaryon made of $V$ field is similar. For our purpose 
we will use, as in eq. (\ref{globalADS}),  the global coordinate system
for the $AdS_5$ part  of the metric and we will take the $Y^{p,q}$ part
as written in eq.  (\ref{alphametric}).  We are interested in the normal
modes of oscillation of the wrapped D3-brane around the solution
corresponding to some fixed worldline in $AdS_5$ and some fixed $\theta$
and $\phi$ on the transverse $S^2$. For such a configuration, the induced
metric on the dibaryon is:
\begin{eqnarray}
L^{-2}ds^2_{ind}=- d\tau^2+\frac{1}{wv}dy^2+\frac{v}{9}d\psi^2+w(d\alpha+fd\psi)^2\,\,,
\end{eqnarray}
where the functions $v(y)$, $w(y)$ and $f(y)$ have been defined in eq. (\ref{qwy}) (in what
follows of this subsection we will take $c=1$).

The fluctuations along the transverse $S^2$ are
the most interesting, since they change the $SU(2)$ quantum
numbers and are most readily compared with the chiral primary
states in the field theory.
Without lost of generality, we consider fluctuations around the
north pole of the $S^2$, i.e. $\theta_0=0$. Instead of using
coordinates $\theta$ and $\phi$, it is convenient to go from polar to Cartesian coordinates:
$\zeta^1=\theta\sin\phi$ and $\zeta^2=\theta\cos\phi$. As a further simplification we
perform a shift in the coordinate $\psi$. The action for the D3-brane is:
\beq
S\,=\,-T_3\,\int d^4\xi \sqrt{-\det g}\,+\,T_3\,\int P[C_{(4)}]\,\,.
\eeq
Let us expand the induced metric $g$ around the static configuration as $g=g_{(0)}+\delta
g$, where  $g_{(0)}$ is the zeroth order contribution. The corresponding expansion for the
action takes the form:
\beq
S\,=\,S_0\,-\,{T_3\over 2}\,\int d^4\xi \sqrt{-\det g_{(0)}}\,\,
{\rm Tr}\,\big[g_{(0)}^{-1}\,\delta g\big]\,+\,T_3\,\int P[C_{(4)}]\,\,,
\eeq
where $S_0=-T_3\,\int d^4\xi \sqrt{-\det g_{(0)}}$.  Note that the determinant of the
induced metric at zeroth order is a constant: 
$\sqrt{-{\rm det}(g_{(0)})}=\frac{1}{3}L^4$.
The five-form field strength is 
\begin{eqnarray}
F_5=(1+*)\, 4\sqrt{{\rm det}(G_{Y^{p,q}})}L^4d\theta\wedge d\phi\wedge
dy\wedge d\psi\wedge d\alpha\,\,.
\end{eqnarray}
Moreover, using that $\sqrt{{\rm det}(G_{Y^{p,q}})}=\frac{1-y}{18}\sin\theta$, we
can choose the four-form Ramond-Ramond field to be
\begin{eqnarray}
C_4=\frac{2}{9}(1-y)L^4(\cos\theta-1)\,d\alpha\wedge dy\wedge d\psi
\wedge d\phi\,,
\end{eqnarray}
which is well defined around the north pole of $S^2$. At  quadratic
order, the four form $C_4$ is 
\begin{eqnarray}
C_4=-\sqrt{-{\rm det}\, g_{(0)} }\,\,\frac{1-y}{3}\epsilon_{ij}\,\zeta^i\,d\zeta^j\wedge
d\alpha\wedge dy\wedge d\psi\,\,.
\end{eqnarray}
The contribution from the Born-Infeld part of the effective action is: 
\beq
{\rm Tr}\,\big[g_{(0)}^{-1}\,\delta g\big]=G_{ij}\;g_{(0)}^{\mu\nu}
\,(\partial_{\mu}\zeta^i\partial_{\nu}\zeta^j)+2g^{\mu\nu}_{(0)}\;
G_{\mu i}\,\partial_{\nu}\zeta^i, 
\eeq
where $G$ is the metric of the background, $i,j$ denote the components of
$G$ along the
$\zeta^{1,2}$ directions and the indices $\mu,\nu$ refer to the directions
of the worldvolume of the cycle. The non-vanishing components of $G$ are:
\beq
G_{ij}=\frac{1-y}{6}L^2\delta_{ij}, \quad G_{\psi i}=
-\frac{1}{2}\bigg(wf^2+\frac{v}{9}\bigg)L^2\epsilon_{ij}\,\zeta^j, \quad 
G_{\alpha i}=-\frac{wf}{2}L^2\epsilon_{ij}\,\zeta^j\,\,. 
\eeq
Using these results one can verify that the effective Lagrangian is
proportional to:
\begin{eqnarray}
\sum_{i}L^2\frac{1-y}{6}g_{(0)}^{\mu\nu}
(\partial_{\mu}\zeta^i\partial_{\nu}\zeta^i)+2g^{\mu\nu}_{(0)}\;G_{\mu
i}\,\partial_{\nu}\zeta^i+\frac{2(1-y)}{3}\epsilon_{ij}
\,\zeta^i\,\partial_{\tau}\zeta^j\,\,.
\end{eqnarray}
The equations of motion for the fluctuation are finally given by
\begin{eqnarray}
{L^2\over 6}\,\partial_{\mu}\bigg((1-y)\,g^{\mu\nu}_{(0)}\,
\partial_{\nu}\zeta^i\,\bigg)+2\partial_{\nu}(g^{\mu\nu}_{(0)}\,G_{\mu
i})-\frac{2(1-y)}{3}\epsilon_{ij}\,\partial_{\tau}\zeta^j=0\,\,.
\end{eqnarray}
 Introducing  $\zeta^{\pm}=\zeta^1\pm i\zeta^2$, the equations of motion
reduce to
\begin{eqnarray}
\bigg(\nabla^2-\frac{1-y}{6}\partial_{\tau}^2\,\bigg) \zeta^{\pm}\pm
i\frac{2(1-y)}{3}\partial_{\tau}\zeta^{\pm}\pm i\partial_{\psi}
\zeta^{\pm}=0\,\,,
\end{eqnarray}
where $\nabla^2$ is the laplacian along the spatial directions of the cycle for the induced
metric $g_{(0)}$. 
The standard strategy to solve this equation is to use separation of
variables as
\begin{eqnarray}
\zeta^{\pm}=\exp(-i\omega\tau)\exp\bigg(i\frac{m}{\ell}\alpha\bigg)\exp(in\psi)
\,Y^{k\pm}_{mn}(y)\,\,.
\end{eqnarray}
Plugging  this ansatz into the equation of motion, we find
\begin{eqnarray}
&&\frac{1}{1-y}\frac{d}{dy}\bigg[(1-y)w(y)v(y)\frac{d}{dy}
Y^{k\pm}_{mn}(y)\bigg]\\
&=&\bigg[\bigg(\frac{9f^2(y)}{v(y)}+\frac{1}{w(y)}\bigg)\frac{m^2}{\ell^2}-
\frac{18f(y)}{v(y)}\frac{m}{\ell}n+\frac{9}{v(y)}n^2-\omega(\omega\pm
4)\pm\frac{6n}{1-y}\bigg]Y^{k\pm}_{mn}(y)\,\,. \nonumber
\end{eqnarray}
The resulting equation has four
regular singularities at $y=y_1,y_2,y_3$ and $\infty$ and is known as 
Heun's equation (for clarity, in what follows we  omit the
indices in $Y$) \cite{heun}:
\be
\frac{d^2}{dy^2}Y^{\pm}+\bigg(\sum_{i=1}^3
\frac{1}{y-y_i}\bigg)\frac{d}{dy}Y^{\pm}+q(y)Y^{\pm}=0,\label{Fuch}
\ee
where, in our case
\begin{eqnarray}
q(y)&=&\frac{2}{{\cal Q}(y)}\bigg[\mu-\frac{y}{4}\omega(\omega\pm
4)-{1\over 2}\,\sum_{i=1}^3\frac{\alpha_i^2
{\cal Q}'(y_i)}{y-y_i}\bigg]\,\,,\rc
\mu&=&\frac{3}{32}(\frac{m}{\ell}+2n)(\frac{m}{\ell}-6n)+\frac{1}{4}\omega(\omega\pm
4)\mp \frac{3n}{2}\,\,, 
\end{eqnarray}
with ${\cal Q}(y)$ being the function defined in eq.(\ref{calq}). 
Now, given that the R-symmetry is dual to the Reeb Killing vector of
$Y^{p,q}$, namely $2\partial/\partial \psi\,-\,
{1\over 3}\,\partial/\partial \alpha$, we can use the R-charge
$Q_R=2n-m/3\ell$ instead of $n$
as  quantum number. The exponents at the regular singularities $y=y_i$ are
then given by
\begin{eqnarray}
\alpha_i=\pm \frac{1}{2}\frac{(1-y_i)(m/\ell+3Q_R\,y_i)}
{ {\cal Q}'(y_i)}.
\end{eqnarray}
The exponents at $y=\infty$ are $-\frac{\omega}{2}$ and
$\frac{\omega}{2}+2$ for $Y^{+}$, while $-\frac{\omega}{2}+2$ and
$\frac{\omega}{2}$ for $Y^{-}$.
We can transform the singularity from $\{y_1,y_2, y_3,\infty\}$ to
$\{0,1,b=\frac{y_1-y_3}{y_1-y_2},\infty\}$ by introducing a new variable $x$, defined as:
\begin{eqnarray}
x=\frac{y-y_1}{y_2-y_1}.
\end{eqnarray}
It is also convenient to substitute
\begin{eqnarray}
Y=x^{|\alpha_1|}(1-x)^{|\alpha_2|}(b-x)^{|\alpha_3|}\,h(x)\,\,,
\end{eqnarray}
which transforms equation (\ref{Fuch}) into the standard form of the
Heun's equation
\be
\frac{d^2}{dx^2}h(x)+\bigg(\frac{C}{x}+\frac{D}{x-1}+
\frac{E}{x-b}\,\bigg)\frac{d}{dx}h(x)+\frac{AB
x-k}{x(x-1)(x-b)}h(x)=0.\label{Heun}
\ee
Here the Heun's parameters are given by
\begin{eqnarray}
A&=&-\frac{\omega}{2}+\sum_{i=1}^3|\alpha_i|\,, \quad\quad
B=\frac{\omega+ 4}{2}+\sum_{i=1}^3|\alpha_i| \,\,,\rc
C&=&1+2|\alpha_1|\,, \quad\quad D=1+2|\alpha_2|\,, \quad\quad
E=1+2|\alpha_3|\,,
\end{eqnarray}
and
\begin{eqnarray}
k&=&(|\alpha_1|+|\alpha_3|)(|\alpha_1|+|\alpha_3|+1)-|\alpha_2|^2\nonumber\\
&+&b\bigg[(|\alpha_1|+|\alpha_2|)(|\alpha_1|+|\alpha_2|+1)-|\alpha_3|^2\bigg]-{\tilde
\mu}\,\,,\rc
{\tilde \mu}&=&-\frac{1}{y_1-y_2}\bigg(\mu-\frac{y_1}{4}\omega(\omega+
4)\bigg)\nonumber\\
&=&\frac{p}{q}\bigg[\frac{1}{6}(1-y_1)\omega(\omega+
4)-\frac{3}{16}Q_R\,\bigg(Q_R+\frac{4m}{3\ell}\bigg)-
\frac{1}{2}\bigg(Q_R+\frac{m}{3\ell}\bigg)\bigg]
\,\,,\label{mu'}\rc
b&=&\frac{1}{2}\bigg(1+\frac{\sqrt{4p^2-3q^2}}{q}\bigg)\,\,.
\end{eqnarray}
We only presented the equation for $Y^{+}$; the
corresponding equation for $Y^{-}$ can be obtained by replacing
$\omega$ with $\omega-4$ and changing the sign of the last term in
(\ref{mu'}).

Now let us discuss the solutions to this differential equation.
For quantum number $Q_R=2N$ (which implies $m=0$), we find all $\alpha_i$
equal to $N/2$. If we set $\omega=3N$, the Heun's parameters
$A$ and $k$ both vanish. The corresponding solution $h(x)$ is
a constant function. Similarly if $\omega=-3N-4$, then $B$ and
$k$ vanish which also implies a constant $h(x)$. The complete
solution of $\zeta^{\pm}$ in these two cases is given by
\begin{eqnarray}
\zeta_1^{\pm}&=&e^{\pm
i(-3N\tau+N\psi)}\prod_{i=1}^{3}(y-y_i)^{N/2}\,\,,\rc
\zeta_2^{\pm}&=&e^{\pm
i((3N+4)\tau+N\psi)}\prod_{i=1}^{3}(y-y_i)^{N/2}\,\,.
\label{zetas}
\end{eqnarray}
These constant solutions represent ground states with fixed quantum
numbers and, since they have the lowest possible dimension for a given
R-charge, they should be identified with the BPS operators. Indeed, in
the solutions (\ref{zetas}) the energy is quantized in units of $3L^{-1}$,
and $3$ is precisely the conformal dimension of ${\cal O}_1$. This
provides a perfect matching of AdS/CFT in this setting.

The situation for quantum numbers $Q_R=N(p\pm q\mp1/3\ell)$ and $m=\pm
N$ is similar to the case we have just discussed. The solutions for
$h(x)$ are constant with
\beq
\omega=\frac{Np}{2}\bigg(3\pm\frac{2p-\sqrt{4p^2-3q^2}}{q}\bigg)\,\,,
\eeq
and
\beq
\omega=-\frac{Np}{2}\bigg(3\pm\frac{2p-\sqrt{4p^2-3q^2}}{q}\bigg)-4.
\eeq
We can see the conformal dimension satisfies
$\Delta=\frac{3}{2}Q_R$. So all these solutions are BPS
fluctuations which should correspond to the operators ${\cal O}_2$
and ${\cal O}_3$.

An interesting comment is in order at this point. Notice that the 
dibaryon excitations should come out with the multiplicities associated
to the SU(2) spin (see Table \ref{chring}) of the ${\cal O}_i$ operators.
However, in order to tackle this problem, we would need to consider at
the same time the fluctuation of the D3-brane probes and the zero-mode
dynamics corresponding to their collective motion along the sphere with
coordinate $\theta$ and $\phi$ (see ref.\cite{ba2} for a similar
discussion in the conifold case). This is an
interesting problem that we leave open.

\setcounter{equation}{0}
\section {Supersymmetric D5-branes in $AdS_5\times Y^{p,q}$}
\label{d5}

In this section we will study the supersymmetric configurations of
D5-branes in the $AdS_5\times Y^{p,q}$ background. First of all, notice
that in this case 
$\Gamma_{\kappa}$ acts on the Killing spinors $\epsilon$ as:

\beq
\Gamma_{\kappa}\,\epsilon\,=\,{i\over 6!\, \sqrt{-g}}\,\,
\epsilon^{\mu_1\cdots\mu_6}\,\gamma_{\mu_1\cdots\mu_6}\,\epsilon^*
\label{Gammakappad5}\,\,,
\eeq
where we have used the relation (\ref{rule}) to translate eq. (\ref{gammakappa}).
The appearance of the complex conjugation on the right-hand side of eq.
(\ref{Gammakappad5}) is crucial in what follows. Indeed, the complex conjugation
does not commute with the projections (\ref{epsilon-project}). Therefore, in order
to construct an additional compatible projection involving the
$\epsilon\to\epsilon^*$ operation we need to include a product of gamma matrices
which anticommutes with both $\Gamma_{12}$ and $\Gamma_{34}$. As in the D3-brane 
case just analyzed, this compatibility requirement between the 
$\Gamma_{\kappa}\,\epsilon = \epsilon$  condition  and (\ref{epsilon-project})
implies a set of differential equations whose solutions, if any, determine the
supersymmetric embeddings we are looking for. 

We will carry out successfully this program only in the case of a D5-brane
extended along a two-dimensional submanifold of $Y^{p,q}$. In analogy with
what happens with the conifold \cite{ba1}, one expects that these kinds of
configurations represent a domain wall in the gauge theory side such
that, when one crosses one of these objects, the gauge groups change and
one passes from an ${\cal N}=1$ superconformal field theory to a
cascading theory with fractional branes. The supergravity dual of this
cascading theory has been obtained in ref.
\cite{hek}. In the remainder of this section we will find the
corresponding configurations of the D5-brane probe. Moreover,  in section
\ref{6} we will find, based on a different set of worldvolume
coordinates, another embedding of this type preserving the same
supersymmetry as the one found in the present section and we will analyze
the effect of adding flux of the worldvolume gauge fields. In section
\ref{6} we will also look at the possibility of having D5-branes wrapped
on a three-dimensional submanifold of 
$Y^{p,q}$. These configurations are not supersymmetric, although we have
been able to  find stable solutions of the equations of motion. The case
in which the D5-brane wraps the entire $Y^{p,q}$ corresponds to the baryon
vertex. In this configuration, studied also in section \ref{6}, the
D5-brane captures the flux of the RR five-form, which acts as a source
for the electric worldvolume gauge field.  We will conclude in section
\ref{6} that this configuration cannot be supersymmetric, in analogy with
what happens in the conifold case\cite{acr}.

\subsection{Domain wall solutions}

We want to find a configuration in which the D5-brane probe wraps a
two-dimensional submanifold of $Y^{p,q}$ and is a codimension one object
in $AdS_5$. Accordingly, let us place the probe at some constant value of
one of the Minkowski coordinates (say $x^3$) and let us extend it along
the radial direction. To describe such an embedding we  choose the following
set of worldvolume coordinates for  a D5-brane probe
\beq
\xi^{\mu}\,=\,(t,x^1,x^2,r,\te,\p)\,\,,
\eeq
and we adopt the following ansatz:
\beq
y=y(\theta,\phi)\,\,,\qquad
\beta\,=\,\beta(\theta,\phi)\,\,,
\label{D5dwansatz}
\eeq
with $x^3$ and $\psi$ constant. The induced Dirac matrices can be computed
straightforwardly from  eq.  (\ref{wvgamma}) with the result:
\bear
&&\gamma_{x^\mu}\,=\,{r\over L}\,\Gamma_{x^\mu}\,\,,\qquad \mu=0,1,2\,\,,\rc\rc
&&\gamma_r\,=\,{L\over r}\,\Gamma_r\,\,,\rc\rc
&&{1\over L}\,\gamma_{\theta}\,=\,-{1\over \sqrt{6}\,H}\,y_\theta\,\Gamma_1\,-\,
{H\over \sqrt{6}}\,\beta_{\theta}\,\Gamma_2\,+\,{\sqrt{1-cy}\over \sqrt{6}}\,
\Gamma_3\,+\,{y\over 3}\,\beta_\theta\,\Gamma_5\,\,,
\qquad\qquad\qquad\qquad\qquad\rc\rc
&&{1\over L}\,\gamma_{\phi} = -{1\over \sqrt{6}\,H}\,y_{\phi}\,\Gamma_1+
{H\over \sqrt{6}}\,(\,c\cos\theta\,-\,\beta_\phi)\,\Gamma_2+{\sqrt{1-cy}\over \sqrt{6}}\,
\sin\theta\,\Gamma_4 \\ \rc
&&\qquad\qquad\qquad+{1\over 3}\big[y\,\beta_\phi +
(1-cy)\cos\theta\big]\,\Gamma_5\,\,.\nonumber
\label{D5gammas}
\eear
From the general expression (\ref{Gammakappad5}) one readily gets that
the kappa symmetry matrix acts on the spinor $\epsilon$ as:
\beq
\Gamma_\kappa\,\epsilon\,=\,{i\over \sqrt{-g}}\,{r^2\over L^2}\,
\Gamma_{x^0x^1x^2r}\,\gamma_{\theta\phi}\,\epsilon^*\,\,.
\eeq
By using the complex conjugate of the projections (\ref{epsilon-project}) one gets:
\beq
{6\over L^2}\,\gamma_{\theta\phi}\,\epsilon^*\,=\,\big[\,
b_I\,+\,b_{15}\,\Gamma_{15}\,+\,b_{35}\,\Gamma_{35}\,+\,b_{13}\,\Gamma_{13}\,\big]\,
\epsilon^*\,\,,
\label{D5gamathetaphi}
\eeq
where the different coefficients are:
\bear
&&b_I\,=\,-i\Big[\,(1-cy)\sin\theta\,+\,c\cos\theta\,y_{\theta}\,+\,
y_\phi\beta_\theta\,-\,y_\theta\beta_\phi\,\Big]\,\,,\rc\rc
&&b_{15}\,=\,-\sqrt{{2\over 3}}\,\,{1\over H}\Big[\,
(1-cy)\cos\theta\,y_\theta\,+\,y\,(\beta_\phi\,y_\theta-\beta_\theta\,y_\phi)\,\Big]\,-\,i
\sqrt{{2\over 3}}\,H\,\cos\theta\,\beta_{\theta}\,\,,\rc\rc
&&b_{35}\,=\,\sqrt{{2\over 3}}\sqrt{1-cy}\,\Big[\,
(1-cy)\cos\theta\,+\,y\beta_\phi\,\Big]\,+\,i
\sqrt{{2\over 3}}\sqrt{1-cy}\,\,y\sin\theta\beta_\theta\,\,,\rc\rc
&&b_{13}\,=\,\sqrt{1-cy}\,\Big[\,{y_\phi\over H}\,-\,H\beta_\theta\,\sin\theta\,\Big]\,+\,
i\sqrt{1-cy}\,\Big[\,{\sin\theta\over H}\,y_{\theta}\,-\,H\,
(c\cos\theta\,-\,\beta_\phi)\,\Big]\,\,.\qquad
\label{D5bs}
\eear
As discussed above,   in this case the action of $\Gamma_{\kappa}$ involves the
complex conjugation, which does not commute with the projections
(\ref{epsilon-project}). Actually, the only term on the right-hand side of
(\ref{D5gamathetaphi}) which is consistent with (\ref{epsilon-project}) is
the one containing $\Gamma_{13}$. Accordingly, we must require:
\beq
b_I\,=\,b_{15}\,=\,b_{35}\,=\,0\,\,.
\eeq
From the vanishing of the imaginary part of $b_{15}$ we get:
\beq
\beta_\theta=0\,\,,
\label{D5beta_theta}
\eeq
while the vanishing of the real part of $b_{15}$ leads to:
\beq
\beta_{\phi}\,=\,-{1-cy\over y}\,\cos\theta\,\,.
\label{betaphi}
\eeq
Notice that $b_{35}$ is zero as a consequence of  equations 
(\ref{D5beta_theta}) and (\ref{betaphi}) which, 
in particular imply that:
\beq
\beta=\beta(\phi)\,\,.
\eeq
Moreover, by using eq. (\ref{D5beta_theta}), the condition $b_I=0$ is equivalent to
\beq
(1-cy)\sin\theta\,+\,(c\cos\theta\,-\,\beta_{\phi})y_\theta\,=\,0\,\,,
\eeq
and plugging the value of $\beta_{\phi}$ from (\ref{betaphi}), one arrives at:
\beq
y_{\theta}\,=\,-(1-cy)\,y\tan\theta\,\,.
\label{D5yeq}
\eeq
In order to implement the kappa symmetry condition at all points of the
worldvolume the phase of $b_{13}$ must be constant. This can be achieved
by requiring that the real part of $b_{13}$ vanishes, which for  
$\beta_\theta=0$   is equivalent to the condition $y_\phi=0$,\,\ie:
\beq
y=y(\theta)\,\,.
\eeq
The equation (\ref{D5yeq}) for $y(\theta)$ is easily integrated, namely:
\beq
{y\over 1-cy}\,=\,k\cos\theta\,\,,
\label{D5ysol}
\eeq
where $k$ is a constant. Moreover, by separating variables  in eq. (\ref{betaphi}), one concludes
that:
\beq
\beta_\phi=m\,\,,
\label{D5phim}
\eeq
where $m$ is a new constant. Plugging (\ref{D5ysol}) into eq.
(\ref{betaphi}) and using the result (\ref{D5phim}) one concludes that
the two constants $m$ and $k$ must be related as:
\beq
km=-1\,\,,
\eeq
which, in particular implies that $k$ and $m$ cannot vanish. 
Thus, the embedding of the D5-brane becomes
\bear
&&\beta=m\phi+\beta_0\,\,,\rc\rc
&&y=-{\cos\theta\over m-c\cos\theta}\,\,.
\label{D5sol}
\eear
Notice that the solution (\ref{D5sol}) is symmetric under the change
$m\to -m$, $\theta\to\pi-\theta$ and $\phi\to 2\pi-\phi$. Thus, from now on we can
assume that $m\ge 0$. 

It is now  straightforward to verify  that the BPS equation are equivalent  to
impose the following condition on the spinor $\epsilon$:
\beq
\Gamma_{x^0x^1x^2r13}\epsilon^*\,=\,\sigma \epsilon\,\,,
\label{d5projector}
\eeq
where $\sigma$ is:
\beq
\sigma\,=\,{\rm sign}\,\Big(\,{\cos\theta\over y}\,\Big)\,=\,
-{\rm sign}\Big(\,m\,-\,c\cos\theta\,\Big)\,\,.
\label{D5sigma-sign}
\eeq
Obviously, the only valid solutions are those which correspond to having a constant sign $\sigma$
along the worldvolume. This always happens for $m/c\ge 1$. In this case the minimal
(maximal) value of $\theta$ is $\theta=0$ ($\theta=\pi$) if 
$|m-c||y_1|>1$ ($|m-c||y_2|>1$). Otherwise the angle $\theta$ must be
restricted to lie in the interval $\theta\in [\theta_{1}, \theta_{2}]$, where
$\theta_{1}$ and $\theta_{2}$ are given by:
\beq
\theta_{i}\,=\,\arccos\Big[{my_i\over cy_i-1}\Big]\,\,,
\qquad\qquad (i=1,2)\,\,.
\eeq
Notice that, similarly to what we obtained in the previous section,
eq.(\ref{D5sol}) implies that the configuration we arrived at does not in
general correspond to a wrapped brane but to a D5-brane that spans a
two-dimensional submanifold with boundaries.

Let us now count the number of supersymmetries preserved by our configuration. In
order to do so we must  convert eq. (\ref{d5projector}) into an algebraic
condition on a constant spinor. With this purpose in mind let us write
the general form of
$\epsilon$ as the sum of the two types of spinors written in eq.
(\ref{chiraladsspinor}), namely:

\beq
e^{{i\over 2}\psi}\,\epsilon\,=\,r^{-{1\over 2}}\,\eta_+\,+\,
r^{{1\over 2}}\,\Big(\,{\bar x^3\over L^2}\,
\Gamma_{rx^3}\,\eta_+\,+\,\eta_-\,\Big)\,+
\,{r^{{1\over 2}}\over L^2}\,x^p\,\Gamma_{rx^p}\,\eta_+\,\,,
\label{generalepsilon}
\eeq
where $\bar x^3$ is the constant value of the coordinate $x^3$ in the embedding and the index
$p$ runs over the set  $\{0,1,2\}$. By substituting eq.  (\ref{generalepsilon}) on both sides of
eq. (\ref{d5projector}), one can get the conditions that $\eta_{+}$  and
$\eta_{-}$ must satisfy. Indeed, let us define the operator ${\cal P}$ as follows:
\beq
{\cal P}\,\epsilon\,\equiv\,i\sigma  e^{i\psi_0}\,\Gamma_{rx^3}\,
\Gamma_{1 3}\,\epsilon^*\,\,.
\eeq
Then, one can check that eq. (\ref{d5projector}) is equivalent to:
\bear
&&{\cal P}\,\eta_+\,=\,\eta_+\,\,,\rc\rc
&&(1\,+\,{\cal P}\,)\,\eta_-\,=\,-{2 \bar x^3\over L^2}\,\Gamma_{rx^3}\,\eta_+\,\,.
\label{d5system}
\eear
As ${\cal P}^2=1$, we can classify the four spinors $\eta_-$ according to their 
${\cal P}$-eigenvalue as: ${\cal P}\,\eta_-^{(\pm)}\,=\,\pm\eta_-^{(\pm)}$.
We can now solve the system (\ref{d5system}) by taking $\eta_+=0$ and 
$\eta_-$ equal to one of the two
spinors $\eta_-^{(-)}$ of negative ${\cal P}$-eigenvalue. Moreover, there are other
two solutions which correspond to taking a spinor $\eta_-^{(+)}$ of positive 
${\cal P}$-eigenvalue and a
spinor $\eta_+$  related to the former as:
\beq
\eta_{+}\,=\,{L^2\over \bar x^3}\,\Gamma_{r x^3}\,\,\eta_-^{(+)}\,\,.
\label{secondspinord5}
\eeq
Notice that, according to the first equation in (\ref{d5system}), the spinor
$\eta_+$ must have positive ${\cal P}$-eigenvalue, in agreement with eq.
(\ref{secondspinord5}). All together this configuration preserves four
supersymmetries, \ie\ one half of the supersymmetries of the background, 
as expected for a domain wall.

\subsection{The calibrating condition}
For any two-dimensional submanifold $L$ of $Y^{p,q}$ one can construct its
three-dimensional cone ${\cal L}\subset CY^{p,q}$.  The holomorphic $(3,0)$ form
$\Omega$ of $CY^{p,q}$ can be naturally used to calibrate such submanifolds.
Indeed,  ${\cal L}$ is called a special Lagrangian submanifold of $CY^{p,q}$ if the
pullback of $\Omega$ to ${\cal L}$ is, up to a constant phase, equal to the volume
form of ${\cal L}$, namely:
\beq
P\big[\,\Omega\,\big]_{{\cal L}}\,=\,e^{i\lambda}\,
{\rm Vol}\,({\cal L})\,\,,
\label{D5calibration}
\eeq
where $\lambda$ is constant on ${\cal L}$. If the cone ${\cal L}$ is special
Lagrangian, its base $L$ is said to be special Legendrian. It has been argued in 
ref. \cite{SY} that the supersymmetric configurations of a D5-brane extended along
a two-dimensional submanifold $L$ of a Sasaki-Einstein space are those for which
${\cal L}$ is special Lagrangian. Let us check that this is indeed the case for the
embeddings (\ref{D5sol}). First of all, we notice that the expression of $\Omega$
written in (\ref{threeform}) can be recast as:
\beq
\Omega\,=\,e^{i\psi}\,r^2\,\Omega_{4}\,\wedge
\big[\,dr+\,i\,{r\over L}\, e^5\,\big]\,\,,
\label{newOmega}
\eeq
where $\Omega_{4}$ is the two-form:
\beq
\Omega_{4}\,=\,{1\over L^2}\,\,\big(\, e^1+ie^2\,\big)\wedge
\big(\, e^3-ie^4\,\big)\,\,.
\label{Omega4}
\eeq
In eqs. (\ref{newOmega}) and (\ref{Omega4})  $e^1$, $\cdots$, $e^5$ are the
vielbein one-forms of (\ref{Ypqvielbein}). Moreover, the volume form of
${\cal L}$ can be written as:
\beq
{\rm Vol}\,({\cal L})\,=\,r^2dr\wedge {\rm Vol}\,( L)\,\,.
\eeq
For our embeddings (\ref{D5sol}) one can check that:
\beq
{\rm Vol}\,( L)\,=\,\,{H\over 6}\,\Big|\,{\cos\theta\over y}\,\Big|\,
\sqrt{1-cy}\,
\Bigg[\,1\,+\,(1-cy)\,{y^2\over H^2}\,\tan^2\theta\,\Bigg]\,d\theta\wedge
d\phi\,\,.
\eeq
It is now straightforward
to verify that our embeddings (\ref{D5sol}) satisfy (\ref{D5calibration}) with
$e^{i\lambda}\,=\,-i\sigma e^{i\psi}$, where $\sigma$ is the constant sign defined
in (\ref{D5sigma-sign}) (recall that in our ansatz (\ref{D5dwansatz}) the
angle $\psi$ is constant). Thus, we conclude that $L$ is special
Legendrian, as claimed. Moreover, one can check that:
\beq
P\big[\,J\,\big]_{{\cal L}}\,=\,0\,\,.
\eeq

\subsection{Energy bound}

Let us consider a generic embedding $y=y(\theta)$, $\beta=\beta(\phi)$ and let
us define the following functions of $\theta$ and $y$
\beq
\Delta_{\theta}\equiv -y(1-cy)\tan\theta\,\,,\qquad
\Delta_{\phi}\equiv-{1-cy\over y}\,\,\cos\theta\,\,.
\eeq
In terms of these functions the BPS equations (\ref{betaphi}) and (\ref{D5yeq})
are simply $y_{\theta}=\Delta_{\theta}$ and
$\beta_{\phi}=\Delta_{\phi}$. We have checked that any solution of this first-order
equations also solves the Euler-Lagrange equations derived from the
Dirac-Born-Infeld lagrangian (\ref{D3lag-den}). Moreover, the 
hamiltonian density 
${\cal H}=\sqrt{-g}$ satisfies a BPS bound as in (\ref{energybound}), where 
${\cal Z}$ is a total derivative. To prove this statement, let us notice that 
${\cal H}$ can be written as:
\bear
&&{\cal H}\,=\,{r^2\over 6}\,{H\over \sqrt{1-cy}}\,
\Big|\,{y\over \cos\theta}\,\Big|\,\sqrt{\Delta_{\phi}^2\,+\,(1-cy)\,
{\cos^2\theta\over y^2 H^2}\,y_{\theta}^2}\,\,\times\rc\rc
&&\qquad\times
\sqrt{(c\cos\theta\,-\,\beta_{\phi})^2\,+\,{\cos^2\theta\over H^2 y^2 (1-cy)}\,
\Delta_{\theta}^2\,+\,{2y^2\over 3H^2}\,(\beta_\phi\,-\,\Delta_\phi)^2}\,\,.
\eear
Let us now rewrite ${\cal H}$ as ${\cal H}=|{\cal Z}|+{\cal S}$, where 
\beq
{\cal Z}\,=\,{r^2\over 6}\,{H\over \sqrt{1-cy}}\,{y\over \cos\theta}\,
\Big[\,{\cos^2 \theta\over y^2 H^2}\,\Delta_{\theta}\,y_{\theta}\,-\,
(c\cos\theta\,-\,\beta_{\phi})\Delta_{\phi}\,\Big]\,\,.
\eeq
One can check that $|{\cal Z}|_{|BPS}\,=\,\sqrt{-g}_{|BPS}$. Moreover, for arbitrary
functions $y=y(\theta)$ and  $\beta=\beta(\phi)$, one can verify that
${\cal Z}$ is a total derivative, namely:
\beq
{\cal Z}\,=\,{\partial \over \partial \theta}\,{\cal Z}^{\theta}\,+\,
{\partial \over \partial \phi}\,{\cal Z}^{\phi}\,\,.
\label{D5calZs}
\eeq
In order to write the explicit expressions of ${\cal Z}^{\theta}$ and
${\cal Z}^{\phi}$, let us define the function $g(y)$ as follows:
\beq
g(y)\,\equiv \, -\int {\sqrt{1-cy}\over H(y)}\,dy\,\,.
\eeq
Then one can verify that eq. (\ref{D5calZs}) is satisfied for ${\cal Z}^{\theta}$ and
${\cal Z}^{\phi}$ given by:
\bear
{\cal Z}^{\theta}&=&{r^2\over 6}\,\sin\theta\,g(y)\,\,,\rc\rc
{\cal Z}^{\phi}&=&{r^2\over 6}\,\Big[\,
-\cos\theta \,g(y)\,\phi\,+\,H(y)\,\sqrt{1-cy}\,\,(c\phi\cos\theta\,-\,\beta)\,\Big]\,\,.
\eear
One can prove that ${\cal H}\ge \big|\,{\cal Z}\,\big|$ is equivalent to:
\bear
&&{\cos^2\theta\over y^2 (1-cy)}\,\Big[\,
\Delta_{\phi}\,\Delta_{\theta}\,+\,(1-cy)\,(c\cos\theta\,-\,\beta_{\phi})\,
y_{\theta}\,\Big]^2\,+\rc\rc
&&\qquad\qquad
{2y^2\over 3}\,\Big[\,\Delta_{\phi}^2\,+\,
{(1-cy)\cos^2\theta\over y^2 H^2}\,y_{\theta}^2\,\Big]\,
[\,\beta_{\phi}\,-\,\Delta_\phi\,]^2\,\ge\, 0\,\,,
\eear
which is always satisfied. 
Moreover, by using that $(c\cos\theta \,-\,\beta_{\phi})_{|BPS}\,=\,\cos\theta/y$, one can
prove that this inequality is saturated precisely when the BPS differential equations are
satisfied.

\setcounter{equation}{0}
\section {Supersymmetric D7-branes in $AdS_5\times Y^{p,q}$}
\label{d7}

For a D7-brane the kappa symmetry matrix (\ref{gammakappa}) takes the form:
\be
\G_k = - \frac{i}{8!\sqrt{-g}}\ep^{\m_1\ldots \m_8}\g_{\m_1\ldots
  \m_8},
\label{D7general-gammak}
\ee
where, again, we have used the rules of eq.(\ref{rule}) to write the
expression of $\G_k$ acting on complex spinors. The D7-branes which fill
the four Minkowski spacetime directions and extend along some holographic
non-compact direction can be potentially used as flavor branes,
\ie\ as branes whose fluctuations can be identified with the dynamical mesons of the
gauge theory. In this section we will find a family of these configurations which
preserve four supersymmetries. In section \ref{6} we will determine another family of
supersymmetric spacetime filling configurations of D7-branes and we will also
demonstrate that there are embeddings in which the D7-brane  wraps the entire 
$Y^{p,q}$ space and preserve two supersymmetries.

\subsection{Spacetime filling D7-brane}

Let us choose a system of worldvolume coordinates motivated by the spacetime filling
character of the configuration that we are trying to find, namely:
\be
\xi=(t,x^1,x^2,x^3,y,\b,\te,\p).
\ee
The ansatz we will adopt for the embedding is:
\be
\psi=\psi(\beta,\phi), \qquad r= r(y,\te).
\label{D7ansatz}
\ee
In this case the general expression  of $\Gamma_{\kappa}$
(eq. (\ref{D7general-gammak})) reduces to:
\beq
\Gamma_{\kappa}\,=\,-i\,{r^4\over L^4\sqrt{-g}}\,
\Gamma_{x^0\cdots x^3}\,\gamma_{y\beta\theta\phi}\,.
\eeq
In order to implement the $\Gamma_{\kappa}\,\epsilon\,=\,\epsilon$ condition we
require that the spinor $\epsilon$ is an eigenvector of the matrix $\Gamma_*$
defined in eq. (\ref{gamma*}). Then, according to eq. (\ref{chiraladsspinor}),  
$\Gamma_*\epsilon=-\epsilon$, \ie\ $\epsilon$ is of the form $\epsilon_-$ and,
therefore, it satisfies:
\beq
\Gamma_{x^0\cdots x^3}\,\epsilon_-\,=\,i\epsilon_-\,\,.
\label{D3chirality}
\eeq
Moreover, 
as $\epsilon_-$ has fixed ten-dimensional chirality, the  condition
(\ref{D3chirality}) implies:
\beq
\Gamma_{r5}\epsilon_-\,=\,-i\epsilon_-\,\,.
\label{Gammar5}
\eeq
By using the projection (\ref{D3chirality}), one immediately arrives at:
\beq
\Gamma_{\kappa}\,\epsilon_-\,=\,{r^4\over L^4\sqrt{-g}}\,
\gamma_{y\beta\theta\phi}\,\epsilon_-\,\,.
\label{D7gamma-epsilon-}
\eeq
The induced gamma matrices appearing on the right-hand side of eq.
(\ref{D7gamma-epsilon-}) are:
\bea
{1\over L}\,\g_y&=&-\frac{1}{\sqrt{6}H}\G_1+\frac{1}{r}r_y\G_r, \nonumber
\\
{1\over L}\,\g_\te &=& \frac{\sqrt{1-c\,y}}{\sqrt{6}}\G_3 +
\frac{1}{r}r_\te
\G_r, \nonumber \\
{1\over L}\,\g_\b&=& -\frac{H}{\sqrt{6}}\G_2+
\frac13\left(\psi_\b+y\right)\G_5,
\nonumber \\
{1\over L}\,\g_\p&=& \frac{cH\cos\te}{\sqrt{6}}\G_2 +\frac{\sqrt{1-c\,
    y}}{\sqrt{6}}\sin\te\G_4 +
\frac13\left(\psi_\p+(1-c\, y)\cos\te\right)\G_5\,\,.
\eea
After using eqs. (\ref{epsilon-project}) and (\ref{Gammar5}),  the action of
$\gamma_{y\beta\theta\phi}$ on
$\epsilon$ can be written as:
\beq
{1\over L^4}\,\gamma_{y\beta\theta\phi}\,\epsilon_-\,=\,
\big[\,d_I\,+\,d_{15}\,\Gamma_{15}\,+\,d_{35}\,\Gamma_{35}\,+\,
d_{13}\Gamma_{13}\,\big]\,\epsilon_-\,\,,
\eeq
where the different coefficients are given by:
\bear
&&d_{I}\,=\,{1-cy\over 36}\,\sin\theta\,+\,{1-cy\over 18}\,\sin\theta\,
(y+\psi_{\beta})\,{r_y\over r}\,-\,{1\over 18}\,
\big[\,(1+c\psi_\beta)\cos\theta+\psi_\phi\,\big]\,{r_\theta\over r}\,\,,
\rc\rc
&&d_{15}\,=\,i\,{1-cy\over 6\sqrt{6}}\,H\,\sin\theta\,
\Big[\,{r_y\over r}\,-\,{y+\psi_\beta\over 3H^2}\,\Big]\,\,,\rc\rc
&&d_{35}\,=\,-i\,{\sqrt{1-cy}\over 6\sqrt{6}}\,\Big[\,\sin\theta\,
{r_\theta\over r}\,+\,{1\over 3}\,
\big(\,(1+c\psi_\beta)\cos\theta\,+\,\psi_{\phi}\,)\,\Big]\,\,,\rc\rc
&&d_{13}\,=\,{{\sqrt{1-cy}\over 18}}\,H\,
\Big[\,\sin\theta\,{y+\psi_\beta\over H^2}\,{r_\theta\over r}\,+\,
\big(\,(1+c\psi_\beta)\,\cos\theta\,+\,\psi_{\phi}\big)\,
{r_y\over r}\,\Big]\,\,.
\eear
As the terms containing the matrices $\Gamma_{15}$, $\Gamma_{35}$ and 
$\Gamma_{13}$ give rise to projections which are not compatible with those 
in eq. (\ref{epsilon-project}), we have to impose that:
\beq
d_{15}\,=\,d_{35}\,=\,d_{13}\,=\,0\,\,.
\eeq
From the vanishing of $d_{15}$ and $d_{35}$ we obtain the following
first-order differential equations
\beq
r_y\,=\,\Lambda_y\,\,, \qquad\qquad
r_\theta\,=\,\Lambda_\theta\,\,,
\label{D7BPSlambda}
\eeq
where we have defined $\Lambda_y$ and $\Lambda_{\theta}$ as:
\bear
&&\Lambda_y\,=\,{r\over 3H^2}\,\big(y+\psi_\beta\big)\,,\rc\rc
&&\Lambda_\theta\,=\,-{r\over 3\sin\theta}\,
\Big[\,(1+c\psi_\beta)\,\cos\theta\,+\,\psi_{\phi}\,\Big]\,\,.
\label{Lambdas}
\eear
Notice that the equations (\ref{D7BPSlambda}) imply that $d_{13}=0$. One can also
check that $r^4\,d_I\,=\,\sqrt{-g}$ if the first-order equations (\ref{D7BPSlambda})
hold and, therefore, one has indeed that $\Gamma_{\kappa}\epsilon_-=\epsilon_-$. 
Thus, any Killing spinor of the type $\epsilon=\epsilon_-$, with 
$\epsilon_-$ as in eq. (\ref{chiraladsspinor}), satisfies the kappa symmetry
condition if the BPS equations  (\ref{D7BPSlambda}) hold. Therefore, these
configurations preserve the four ordinary supersymmetries of the background and, as
a consequence, they are 1/8 supersymmetric.

\subsection{Integration of the first-order equations}
Let us now obtain the general solution of the system (\ref{D7BPSlambda}). Our first
observation is that, according to (\ref{D7ansatz}), the only dependence on the
coordinates $\beta$ and $\phi$ appearing in eqs. (\ref{D7BPSlambda}) and
(\ref{Lambdas}) comes from the derivatives of $\psi$. Therefore, for consistency
with the assumed dependence of the functions of the ansatz
(\ref{D7ansatz}),
$\psi_\phi$ and $\psi_\beta$ must be constants. Thus, let us write:
\beq
\psi_\phi\,=\,n_1\,\,,\qquad\qquad
\psi_\beta\,=\,n_2\,\,,
\eeq
which can be  trivially integrated, namely:
\beq
\psi\,=\,n_1\,\phi\,+\,n_2\,\beta\,+\,{\rm constant}\,\,.
\label{D7psi-solution}
\eeq
It is now easy to obtain the function $r(\theta, y)$. 
The equations to integrate are:
\beq
r_y\,=\,{r\over 3H^2}\,(y+n_2)\,\,,\qquad
r_\theta\,=\,-{r\over 3\sin\theta}\,
\Big[\,(1+cn_2)\cos\theta\,\,+\,n_1\,\Big]\,\,.
\label{D7-rtheta-ry}
\eeq
Let us first integrate the equation for $r_\theta$ in 
(\ref{D7-rtheta-ry}). We get:
\beq
r(y,\te)\,=\,{A(y)\over \Big[
\sin {\theta\over 2}\Big]^{{1+n_1+cn_2\over 3}}\,\,
\Big[\cos {\theta\over 2}\Big]^{{1-n_1+cn_2\over 3}}}\,\,,
\eeq
with $A(y)$ a function of $y$ to be determined. Plugging this result in the
equation for $r_y$ in (\ref{D7-rtheta-ry}), we get the following equation for $A$:
\beq
{1\over A}\,{dA\over dy}\,=\,{1\over 3}\,{y+n_2\over H^2}\,\,,
\eeq
which can be integrated immediately, namely:
\beq
A^3(y)\,=\,C\,\Big [f_1(y)\Big]^{n_2}\,f_2(y)\,\,,
\eeq
with $C$ a constant and $f_1(y)$ and $f_2(y)$ being the functions defined in
(\ref{fs}). Then, we can write $r(y,\te)$ as:
\beq
r^3(y,\te)\,=\,C{\Big [f_1(y)\Big]^{n_2}\,f_2(y)
\over \Big[
\sin {\theta\over 2}\Big]^{1+n_1+cn_2}\,\,
\Big[\cos {\theta\over 2}\Big]^{1-n_1+cn_2}}\,\,.
\label{D7r-solution}
\eeq
Several comments concerning the solution displayed in eqs.
(\ref{D7psi-solution}) and (\ref{D7r-solution}) are in order. First of
all, after a suitable change of variable it is easy to verify that for
$c=0$ one recovers from (\ref{D7psi-solution}) and (\ref{D7r-solution})
the family of D7-brane embeddings in
$AdS_5\times T^{1,1}$ found in ref. \cite{acr}. Secondly, the function 
$r(y,\te)$ in (\ref{D7r-solution}) always diverges for some particular
values of $\theta$ and $y$, which means that the probe always extends
infinitely in the holographic direction. Moreover, for some particular
values of $n_1$ and $n_2$ there is a minimal value of the coordinate $r$,
which depends on the integration constant $C$. This fact is important
when one tries to use these D7-brane configurations as flavor branes,
since this minimal value of $r$ provides us with an energy scale, which
is naturally identified with the mass of the dynamical quarks added to the
gauge theory. It is also interesting to obtain the form of the solution
written in eqs. (\ref{D7psi-solution}) and (\ref{D7r-solution}) 
in terms of the complex variables $z_i$ defined in (\ref{complexzs}).
After a simple calculation one can verify that this solution can be
written as a polynomial  equation of the form:
\beq
z_1^{m_1}\,z_2^{m_2}\,z_3^{m_3}\,=\,{\rm constant}\,\,,
\label{D7polynomial}
\eeq
where the $m_i$'s are constants and $m_3\not= 0$.\footnote{
It is natural to expect a condition of the form $f(z_1,z_2,z_3)=0$, where
$f$ is a general holomorphic function of its arguments. However, in order
to be able to solve the  problem analytically we started from a
restrictive ansatz (\ref{D7ansatz}) that, not surprisingly, leads to a
particular case of the expected answer.} The relation
between the $m_i$'s of (\ref{D7polynomial}) and the $n_i$'s of eqs. 
(\ref{D7psi-solution}) and (\ref{D7r-solution}) is:
\beq
n_1\,=\,{m_1\over m_3}\,\,,\qquad
n_2\,=\,{m_2\over m_3}\,\,.
\eeq
Notice that when $n_2=m_2=0$ the dependence on $\beta$ disappears and the
configuration is reminiscent of its analog in the conifold case
\cite{acr}. When $n_2\not= 0$ the D7-brane winds infinitely the
$\psi$-circle.

\subsection{Energy bound}

As it happened in the case of D3- and D5-branes, one can verify that any solution of the first-order equations (\ref{D7BPSlambda}) also solves the equations of motion. We are now going to check that there exists a bound
for the energy which is saturated
by the solutions of the first-order equations  (\ref{D7BPSlambda}). Indeed, let 
$r(y, \theta)$ and $\psi(\beta,\phi)$ be arbitrary functions. The hamiltonian
density ${\cal H}=\sqrt{-g}$ in this case can be written as:
\beq
{\cal H}\,=\,{r^2\over 6}\,\sin\theta\,
\sqrt{\Bigg(r_\theta^2\,+\,(1-cy)\,\Big[H^2\,r_y^2\,+\,{r^2\over 6}\Big]\Bigg)\,
\Bigg(\Lambda_\theta^2\,+\,(1-cy)\,\Big[H^2\,\Lambda_y^2\,+\,{r^2\over 6}\Big]\Bigg)}\,\,,
\eeq
where $\Lambda_y$ and $\Lambda_{\theta}$ are the functions displayed in 
eq. (\ref{Lambdas}). 
Let us rewrite this function ${\cal H}$  
as ${\cal Z}+{\cal S}$, where  ${\cal Z}$ is given
by:
\beq
{\cal Z}\,=\,{r^2\over 6}\,\sin\theta\,\Bigg[\,
r_\theta\Lambda_\theta\,+\,(1-cy)\,\Big(H^2\, r_y\,\Lambda_y\,+\,{r^2\over 6}
\Big)\Bigg] ~.
\label{D7calZ}
\eeq
One can prove that ${\cal Z}$ is a total derivative:
\beq
{\cal Z}\,=\,\partial_{\theta}\,{\cal Z}^\theta\,+\,\partial_y\,{\cal Z}^y\,\,,
\eeq
where ${\cal Z}^\theta$ and ${\cal Z}^y$ are:
\bear
&&{\cal Z}^\theta\,=\,-{r^4\over
72}\,\Big[\,\psi_\phi\,+\,(1+c\psi_\beta)\,\cos\theta\,\Big]\,\,,\rc\rc 
&&{\cal Z}^y\,=\,{r^4\over 72}\,(1-cy)\,(y+\psi_\beta)\sin\theta\,\,.
\eear
Moreover, when ${\cal Z}$ is given by (\ref{D7calZ}), one can demonstrate the bound
(\ref{energybound}). Actually, one can show that the condition 
${\cal H}\ge |{\cal Z}|$ is equivalent to the inequality:
\beq
(r_\theta-\Lambda_\theta)^2\,+\,H^2\,(1-cy)\,(r_y-\Lambda_y)^2\,+\,
{H^2\over r^2}\,(r_\theta\,\Lambda_y-r_y\,\Lambda_\theta)^2\,\ge 0\,\,,
\eeq
which is always satisfied and is saturated precisely when the BPS equations 
(\ref{D7BPSlambda}) are satisfied. Notice also that ${\cal Z}_{|BPS}$ is positive.

\setcounter{equation}{0}
\section{Other interesting possibilities}
\label{6}
Let us now look at some other configurations of different branes and cycles not 
considered so far. We  first consider D3-branes extended along one of the
Minkowski coordinates and along a two-dimensional submanifold of
$Y^{p,q}$. These configurations represent ``fat" strings  from the point
of view of the gauge theory. We  verify in subsection \ref{fat} that
an embedding of this type breaks completely the supersymmetry, although
there exist stable non-supersymmetric fat strings.  In subsection
\ref{MoreDW} we  find a new configuration of a D5-brane wrapping a
two-dimensional submanifold, whereas in subsection \ref{D5flux} we 
add  worldvolume flux to the domain wall solutions of section \ref{d5}. 
In subsection \ref{d5-3cycle} we  consider the possibility of having
D5-branes wrapping a three-cycle. We  show that such embeddings
cannot be supersymmetric, even though stable solutions of the equations
of motion with these characteristics do exist. In subsection
\ref{baryonvertex} we  analyze the baryon vertex configuration (a
D5-brane wrapping the entire $Y^{p,q}$ space) and we  verify that
such embedding breaks supersymmetry completely. In subsection 
\ref{MoreD7} we  explore the existence of spacetime filling
supersymmetric configurations of D7-branes by using a set of worldvolume
coordinates different from those used in section \ref{d7}. Finally, in
subsection \ref{D7Ypq} we  show  that a  D7-brane can wrap the whole
$Y^{p,q}$ space and preserve some fraction of supersymmetry. 

\subsection{D3-branes  on a two-submanifold}
\label{fat}

Let us take a D3-brane which is extended along one of the spatial directions of the worldvolume of
the D3-branes of the background (say $x^1$) and wraps a two-dimensional cycle. The worldvolume
coordinates we will take are
\beq
\xi^{\mu}\,=\,(x^0, x^1, \theta, \phi)\,\,,
\eeq
and we will look for embeddings with $x^2$, $x^3$, $r$ and $\psi$ constant and with
\beq
y\,=\,y(\theta,\phi)\,\,,\quad\qquad
\beta\,=\,\beta(\theta,\phi)\,\,.
\eeq
In this case the kappa symmetry matrix acts on $\epsilon$ as:
\beq
\Gamma_{\kappa}\,\epsilon\,=\,-{i\over \sqrt{-g}}\,{r^2\over L^2}\,
\Gamma_{x^0x^1}\,\gamma_{\theta\phi}\,\epsilon\,\,.
\eeq
The expressions of $\gamma_{\theta}$ and $\gamma_{\phi}$ are just those given in eq.
(\ref{D5gammas}). Moreover, $\gamma_{\theta\phi}\,\epsilon$ can be obtained by taking the
complex conjugate of  eq. (\ref{D5gamathetaphi}):
\beq
{6\over L^2}\,\gamma_{\theta\phi}\,\epsilon\,=\,\big[\,
b_I^*\,+\,b_{15}^*\,\Gamma_{15}\,+\,b_{35}^*\,\Gamma_{35}\,+\,b_{13}^*\,\Gamma_{13}\,\big]\,
\epsilon\,\,,
\eeq
where the $b$'s are given in eq. (\ref{D5bs}). Since now the complex conjugation does not act on
the spinor $\epsilon$, the only possible projection compatible with those of the background is the
one originated from the term with the unit matrix in the previous expression. Then, we must
require:
\beq
b_{15}=b_{35}=b_{13}=0\,\,.
\eeq
The conditions $b_{15}=0$ and $b_{35}=0$ are equivalent and give rise to eqs. (\ref{D5beta_theta})
and (\ref{betaphi}), which can be integrated as in eq. (\ref{D5sol}). Moreover, the condition
$b_{13}=0$ leads to the equation:
\beq
{y\over H^2}\,y_{\theta}\,=\,\cot\theta\,\,.
\eeq
The integration of this equation can be straightforwardly performed in terms of the function
$f_2(y)$ defined in eq. (\ref{fs}) and can be written as:
\beq
{1\over \sqrt{a-3y^2+2cy^3}}\,=\,k\sin\theta\,\,,
\eeq
with $k$ being a constant of integration, which should be related to the constant $m$ in eq.
(\ref{D5sol}). However, the dependence of $y$ on $\theta$ written in the last equation does not
seem to be compatible with the one of eq. (\ref{D5sol}) (even for $c=0$). Thus, we conclude that
there is no solution for the kappa symmetry condition in this case. 

If we forget about the requirement of supersymmetry it is not difficult to find solutions of the
Euler-Lagrange equations of  motion of the D3-brane probe. Indeed, up to irrelevant global
factors, the lagrangian for the D3-brane considered here is the same as the one corresponding to a
D5-brane extended along a two-dimensional submanifold of $Y^{p,q}$. Thus, 
the embeddings written in eq. (\ref{D5sol}) are stable solutions of the
equations of motion of the D3-brane which represent  a ``fat string" from
the gauge theory point of view.

\subsection{More D5-branes wrapped on a two-cycle}
\label{MoreDW}

Let us consider a D5-brane wrapped on a two-cycle and let us choose the following set of
worldvolume coordinates: $\xi^{\mu}\,=\,(x^0,x^1,x^2,r,\theta,y)$. The embeddings we shall
consider have $x^3$ and $\psi$ constant and $\phi=\phi(\theta, y)$, $\beta=\beta(\theta, y)$. For
this case, one has:
\beq
\Gamma_{\kappa}\,\epsilon\,=\,{i\over \sqrt{-g}}\,{r^2\over L^2}\,\,
\Gamma_{x^0x^1x^2r}\,\,\gamma_{\theta y}\,\epsilon^*\,.
\eeq
The induced gamma matrices are:
\bear
&&{1\over L}\gamma_{\theta}\,=\,{H\over
\sqrt{6}}\,\Big(\,-\beta_{\theta}\,+\,c\cos\theta\phi_{\theta}\,\Big)\Gamma_2\,+\,
\sqrt{{1-cy\over 6}}\,\Big(\,\Gamma_3\,+\,\sin\theta\phi_{\theta}\Gamma_4\,\Big)\,+\,\rc\rc
&&\qquad\qquad+{1\over 3}\,\Big(\,y\beta_{\theta}\,+\,(1-cy)\cos\theta\,\phi_{\theta}\,\Big)\,
\Gamma_5\,\,,\rc\rc
&&{1\over L}\,\gamma_{y}\,=\,-{1\over \sqrt{6} H}\,\Gamma_1\,+\,{H\over \sqrt{6}}\,
\Big(\,-\beta_y\,+\,c\cos\theta\phi_y\,\Big)\,\Gamma_2\,+\,
\sqrt{{1-cy\over 6}}\,\sin\theta\phi_y\,\Gamma_4\,+\,\rc\rc
&&\qquad\qquad+{1\over 3}\,\Big(\,y\beta_y+(1-cy)\cos\theta\phi_y\,\Big)\,\Gamma_5\,\,.
\eear
Then, one has

\beq
{6\over L^2}\,\gamma_{\theta y}\,\,\epsilon^*\,=\,
\Big(\,f_I\,+\,f_{15}\Gamma_{15}\,+\,f_{35}\Gamma_{35}\,+\,
f_{13}\Gamma_{13}\,\Big)\,\epsilon^*\,\,,
\eeq
where the different coefficients are given by:
\bear
&&f_I\,=\,-i\Big(\,(1-cy)\,\sin\theta\,\phi_y\,-\,c\cos\theta\phi_\theta\,
+\,\beta_\theta\,\Big)\,\,,\rc\rc
&&f_{15}\,=\,\sqrt{{2\over 3}}\,{1\over H}\,\Big(\,y\beta_\theta\,+\,
(1-cy)\cos\theta\phi_\theta\,\Big)\,+\,i\sqrt{{2\over 3}}\,H\,\cos\theta\,
\Big(\,\beta_y\,\phi_\theta\,-\,\beta_\theta\,\phi_y\,\Big)\,\,,\\\rc
&&f_{35}\,=\,\sqrt{{2\over 3}}\,\sqrt{1-cy}\,\Big[ \Big(\,y\beta_y\,+\,
(1-cy)\cos\theta\phi_y\,\Big)\,-\,i\,y\,\sin\theta\,
\Big(\,\beta_y\,\phi_\theta\,-\,\beta_\theta\,\phi_y\,\Big) \Big]\,\,,\rc\rc
&&f_{13}\,=\,\sqrt{1-cy}\, \Big[ \Big(\,{1\over H}\,+\,H\sin\theta\,
(\,\beta_y\,\phi_\theta\,-\,\beta_\theta\,\phi_y\,)\Big)
- i\,\Big( {\sin\theta\over H}\,\phi_\theta - H
(\beta_y-c\cos\theta\phi_y)\Big) \Big]\,\,.\nonumber
\eear
The BPS conditions  in this case are the following:
\beq
f_I\,=\,f_{15}\,=\,f_{35}\,=\,0\,\,.
\eeq
From the vanishing of $f_I$ we get the equation:
\beq
\beta_{\theta}\,+\,(1-cy)\sin\theta\,\phi_y\,-
\,c\cos\theta\phi_\theta\,=\,0\,\,.
\label{fIeq}
\eeq
Moreover, the vanishing of $f_{15}$ and $f_{35}$ is equivalent to the equations:
\bear
&&y\beta_{\theta}\,+\,(1-cy)\,\cos\theta\phi_\theta\,=\,0\,\,,\rc\rc
&&y\beta_y\,+\,(1-cy)\cos\theta\,\phi_y\,=\,0\,\,,\rc\rc
&&\beta_y\,\phi_\theta-\beta_\theta\,\phi_y\,=\,0\,\,.
\label{f5eqs}
\eear
Notice that this system of equations is redundant, \ie\ the first two equations are equivalent
if one uses the last one. 
Substituting the value of $\beta_\theta$ as given by  the first equation in 
(\ref{f5eqs}) into  (\ref{fIeq}), one can get a partial differential equation which only
involves derivatives of 
$\phi$, namely:
\beq
\cot\theta\,\phi_{\theta}\,-\,y(1-cy)\,\phi_y\,=\,0\,\,.
\label{D5phi}
\eeq
By using in (\ref{D5phi}) the last equation in (\ref{f5eqs}), one gets:
\beq
\cot\theta\,\beta_{\theta}\,-\,y(1-cy)\,\beta_y\,=\,0\,\,.
\label{D5beta}
\eeq
Eqs. (\ref{D5phi}) and (\ref{D5beta}) can be easily integrated by the method of separation of
variables. One gets
\bear
&&\phi\,=\,A\,\Bigg[{y\over (1-cy)\cos\theta}\,\Bigg]^{\alpha}\,+\,\phi^0\,\,,\rc\rc
&&\beta\,=\,{\alpha \over 1-\alpha}\,A\,\Bigg[{y\over
(1-cy)\cos\theta}\,\Bigg]^{\alpha-1}\,+\,\beta^0\,\,,
\eear
where $A$, $\alpha$, $\phi^0$ and $\beta^0$  are constants of integration and we have used 
eq. (\ref{f5eqs}) to relate the integration constants of $\phi$ and $\beta$. However, in order
to implement the condition $\Gamma_{\kappa}\,\epsilon=\epsilon$, one must require 
the vanishing of the
imaginary part of $f_{13}$. This only happens if $\phi$ and $\beta$ are constant,
\ie\ when $A=0$ in the above solution. One can check that this configuration satisfies the
equation of motion.

\subsection{D5-branes  on a two-submanifold with flux}
\label{D5flux}

We now analyze the effect of adding flux of the worldvolume gauge field $F$
to the configurations of section 4 \footnote{A nice discussion of supersymmetric
configurations with nonzero gauge field strengths by means of kappa symmetry
can be found in Ref.\cite{mmms}.}. Notice that  we now have a non-zero contribution from the Wess-Zumino term of the action, which is of the form:
\beq
{\cal L}_{WZ}\,=\,P[\,C^{(4)}\,]\wedge F\,\,.
\eeq
Let us suppose that we switch on a worldvolume gauge field along the angular directions
$(\theta,\phi)$. We will adopt the ansatz:
\beq
F_{\theta\phi}\,=\,q\,K(\theta,\phi)\,\,,
\label{wvflux}
\eeq
where $q$ is a constant and $K(\theta,\phi)$ a function to be determined. The relevant components
of $P[\,C^{(4)}\,]$ are
\beq
P[\,C^{(4)}\,]_{x^0x^1x^2r}\,=\,h^{-1}\,{\partial x^3\over \partial r}\,\,,
\eeq
where $h=L^4/r^4$.
It is clear from the above expression of ${\cal L}_{WZ}$ that a nonvanishing value of  $q$
induces a dependence of $x^3$ on $r$. In what follows we will assume that $x^3=x^3(r)$, \ie\ that 
$x^3$ only depends on $r$. Let us assume that the angular embedding satisfies the same equations 
as in the case of zero flux. The Lagrangian density in this case is given by:
\beq
{\cal L}\,=\,-h^{-{1\over 2}}\,\sqrt{1+h^{-1}\,(x')^2}\,
\sqrt{g_{\theta\theta}g_{\phi\phi}\,+\,q^2\,K^2}\,+\,q\,h^{-1}x'K\,\,,
\eeq
where $g_{\theta\theta}$ and $g_{\phi\phi}$ are elements of the induced metric, we have denoted
$x^3$ simply by $x$ and the prime denotes derivative with respect to $r$. The equation of motion
of $x$ is:
\beq
-{\sqrt{g_{\theta\theta}g_{\phi\phi}\,+\,q^2\,K^2}\over \sqrt{1+h^{-1}\,(x')^2}}\,
h^{-{3\over 2}}\,x'\,+\,q\,h^{-1}\,K\,=\,{\rm constant}\,\,.
\eeq
Taking the constant on the right-hand side of the above equation equal to zero, we get the
following solution for $x'$:
\beq
x'(r)\,=\,q\,h^{{1\over 2}}\,
{K(\theta,\phi)\over \sqrt{g_{\theta\theta}g_{\phi\phi}}}\,\,.
\eeq
Notice that the left-hand side of the above equation depends only on $r$, whereas the right-hand
side can depend on the angles $(\theta,\phi)$. For consistency the dependence of 
$K(\theta,\phi)$ and $\sqrt{g_{\theta\theta}g_{\phi\phi}}$ on $(\theta,\phi)$ must be the same.
Without lost of generality let us take $K(\theta,\phi)$ to be:
\beq
L^2\,K(\theta,\phi)\,=\,\sqrt{g_{\theta\theta}g_{\phi\phi}}\,\,,
\eeq
where the factor $L^2$ has been introduced for convenience. Using this form of $K$, the
differential equation which determines the dependence of $x^3$ on $r$
becomes:
\beq
x'(r)\,=\,{q\over r^2}\,\,,
\eeq
which can be immediately integrated, namely:
\beq
x(r)\,=\,\bar x^3\,-\,{q\over r}\,\,.
\label{x(r)}
\eeq
Moreover, the expression of $K$ can be obtained by computing the induced metric along the angular
directions.  It takes the form:
\beq
K(\theta)\,=\,\sigma\,{\sqrt{1-cy}\over 6 H(y)}\,\Big[\,H^2(y)\,+\,(1-cy)y^2\,\tan^2\theta\,\Big]\,
{\cos\theta\over y}\,\,,
\eeq
where $y=y(\theta)$ is the function obtained in section 4 and
$\sigma\,=\,{\rm sign} \Big(\cos\theta/y\Big)$. Actually, notice that $K$ only depends on the angle
$\theta$ and it is independent of $\phi$.

We are now going to verify that the configuration just found is
supersymmetric. The expression of $\Gamma_{\kappa}$ in this case has an additional term due to the
worldvolume gauge field. Actually, it is straightforward to check that in the present case
\beq
\Gamma_{\kappa}\,\epsilon\,=\,{i\over 
\sqrt{-\det (g+F)}}\,\,{r^3\over L^3}\,\Gamma_{x^0x^1x^2}\,
\Bigg[\,\gamma_r\,\gamma_{\theta\phi}\,\epsilon^*\,
-\,\gamma_r\,F_{\theta\phi}\,\epsilon\,\Bigg]\,\,.
\eeq
Notice that $\gamma_r$ is given by:
\beq
\gamma_r\,=\,{L\over r}\,\big(\,\Gamma_r\,+\,{r^2\over L^2}\,x'\,
\Gamma_{x^3}\,\big)\,\,.
\eeq
For the angular embeddings we are considering it is easy to prove from the results of section 4
that:
\beq
\gamma_{\theta\phi}\,\epsilon^*\,=\,-i \sigma L^2 K(\theta)\, 
\Gamma_{13}\,\epsilon^*\,\,.
\eeq
By using this result and the value of $F_{\theta\phi}$ (eq. (\ref{wvflux})), one easily verifies
that:
\beq
\Gamma_{\kappa}\,\epsilon\,=\,-{i\over 1+{q^2\over L^4}}\,\,
\Gamma_{x^0x^1x^2 r}\Big[\,
i\sigma\Gamma_{13}\,\epsilon^*\,+\,
{q\over L^2}\, i\sigma\,\Gamma_{r x^3}\,\Gamma_{13}\,\epsilon^* +
{q\over L^2}\,\epsilon\,+\,{q^2\over L^4}\,\Gamma_{rx^3}\,\epsilon\,\Big]\,\,.
\eeq
By using the explicit dependence of $x$ on $r$ (eq. (\ref{x(r)})), one can write the Killing
spinor $\epsilon$ evaluated on the worldvolume as:
\beq
e^{{i\over 2}\psi}\,\epsilon\,=\,r^{-{1\over 2}}\,\Big(\,1\,-\,{ q\over L^2}\,\Gamma_{rx^3}\Big)
\eta_+\,+\,
r^{{1\over 2}}\,\Big(\,{\bar x^3\over L^2}\,
\Gamma_{rx^3}\,\eta_+\,+\,\eta_-\,\Big)\,+
\,{r^{{1\over 2}}\over L^2}\,x^p\,\Gamma_{rx^p}\,\eta_+\,\,,
\label{epsilonflux}
\eeq
where the constant spinors $\eta_{\pm}$ are the ones defined in eq. (\ref{etamasmenos}).
Remarkably, one finds that the condition $\Gamma_{\kappa}\epsilon=\epsilon$ is verified if
$\eta_{+}$ and $\eta_{-}$ satisfy the same system (\ref{d5system}) as is the case of zero
flux. 

\subsection{D5-branes wrapped on a three-cycle}
\label{d5-3cycle}

We will now try to find supersymmetric configurations of D5-branes wrapping a three cycle of the 
$Y^{p,q}$ space. Let us choose the following set of worldvolume coordinates 
$\xi^{\mu}=(x^0, x^1, x^2, y, \beta, \psi)$  and consider an embedding with $x^3$ and $r$ constant, 
 $\theta=\theta(y,\beta)$ and $\phi=\phi(y,\beta)$. In this case:
\beq
\Gamma_{\kappa}\,\epsilon\,=\,{i\over \sqrt{-g}}\,\,
{r^3\over L^3}\,\,\Gamma_{x^0x^1x^2}\,\gamma_{y\beta\psi}\,
\epsilon^*\,\,.
\eeq
The value of $\gamma_{y\beta\psi}\,\epsilon^*$ can be obtained by taking the complex conjugate 
of eq. (\ref{HEKcs}). As $c_1=c_3=0$ when $\theta_{\psi}=\phi_{\psi}=0$, we can write:
\beq
{i\over L^3}\,\gamma_{y\beta\psi}\,\epsilon^*\,=\,
\big[\,c_5^*\,\Gamma_5\,+\,
c_{135}^*\,\Gamma_{135}\,\big]\,\epsilon^*\,\,.
\eeq
The only possible BPS condition compatible with the projections satisfied by $\epsilon$ is
$c_5=0$, which leads to a projection of the type 
\beq
\Gamma_{x^0x^1x^2}\Gamma_{135}\,\epsilon^*\,=\,\lambda \epsilon\,\,,
\eeq
where $\lambda$ is a phase.
Notice that, however, as the spinor $\epsilon$ contains a factor $e^{-{i\over 2}\psi}$, the
two sides of the above  equation depend differently on $\psi$ due to the complex
conjugation appearing on the left-hand side ($\lambda$ does not depend on $\psi$). Thus,
these configurations cannot be supersymmetric. We could try to use another set of
worldvolume coordinates, in particular one which does not include $\psi$. After some
calculation one can check that there is no consistent solution. 

For the ansatz considered above the lagrangian density of the D5-brane is, up to irrelevant
factors, the same as the one obtained in subsection \ref{hekconst} 
 for a D3-brane wrapping a  three-dimensional submanifold of $Y^{p,q}$. Therefore
any solution of the  first-order equations (\ref{HEKBPS}) gives rise to an embedding
of a D5-brane which  solves the equations of motion and saturates an energy bound. This
last fact implies that the D5-brane configuration is stable, in spite of
the fact that it is not supersymmetric.  

\subsection{The baryon vertex}
\label{baryonvertex}

If a D5-brane wraps the whole $Y^{p,q}$ space, the flux of the Ramond-Ramond five form 
$F^{(5)}$ that it captures acts as a source for the electric worldvolume gauge field which, in
turn, gives rise  to a bundle of fundamental strings emanating from the D5-brane. This is
the basic argument of  Witten's construction of the baryon vertex
\cite{ba0}, which we will explore in detail now. In this case the probe
action  must include the worldvolume gauge field $F$ in both the
Born-Infeld and Wess-Zumino terms. It takes the form:
\beq
S\,=\,-T_5\,\int d^6\xi\,\sqrt{-\det (g+F)}\,+\,
T_5\int d^6\xi \,\,\,A\wedge F^{(5)}\,\,,
\label{baryonaction}
\eeq
where $T_5$ is the tension of the D5-brane and $A$ is the one-form potential for $F$ (
$F=dA$). In order to analyze the contribution of the Wess-Zumino term in
(\ref{baryonaction})  let us rewrite the expression (\ref{F5}) of $F^{(5)}$ as:
\beq
F^{(5)}\,=\,{L^4\over 27}\,\,(1-cy)\,\sin\theta\,
dy\wedge d\beta\wedge d\theta\wedge d\phi \wedge d\psi\,+\,
{\rm Hodge}\,\,\, {\rm dual}\,\,,
\label{baryonF5}
\eeq
where, for simplicity we are taking the string coupling constant $g_s$ equal to one. Let us
also choose the following set of worldvolume coordinates:
\beq
\xi^{\mu}\,=\,(x^0,y,\beta,\theta,\phi,\psi)\,\,.
\label{baryon-coordinates}
\eeq
It is clear from the expressions of $F^{(5)}$ in (\ref{baryonF5}) and  of the
Wess-Zumino term in (\ref{baryonaction}) that, for consistency, we must turn on the time
component of the field $A$. Actually, we will adopt the following ansatz:
\beq
r\,=\,r(y)\,\,,\qquad\qquad
A_0\,=\,A_0(y)\,\,.
\label{baryon-ansatz}
\eeq
The action (\ref{baryonaction}) for such a configuration can be written as:
\beq
S\,=\,{T_5 L^4\over 108}\,\,V_4\,\,\int dx^0 dy\,\,
{\cal L}_{eff}\,\,,
\label{Leff}
\eeq
where the volume $V_4$ is :
\beq
V_4\,=\,6\int d\alpha\, d\psi\, d\phi\,
d\theta\sin\theta\,=\,96\pi^3\,\ell\,,
\eeq
and the effective lagrangian density ${\cal L}_{eff}$  is given by:
\beq
{\cal L}_{eff}\,=\,(1-cy)\,\Bigg[\,-H\,
\sqrt{{r^2\over H^2}\,+\,6\,(r')^2\,-\,6\,(F_{x^0 y})^2}\,+\,4A_0\,\Bigg]\,\,.
\eeq
Notice that, for our ansatz (\ref{baryon-ansatz}), the electric field 
is $F_{x^0 y}=-\partial_y A_0$. 
Let us now introduce the displacement field, defined as:
\beq
D(y) \equiv {\partial {\cal L}_{eff}\over \partial F_{x^0y}}\,=\,
{6(1-cy)HF_{x^0 y}\over 
\sqrt{{r^2\over H^2}\,+\,6\,(r')^2\,-\,6\,(F_{x^0 y})^2}}\,\,.
\label{D}
\eeq
From the equations of motion of the system it is straightforward to determine $D(y)$.
Indeed, the variation of $S$ with respect to $A_0$ gives rise to the Gauss' law:
\beq
{dD(y)\over dy}\,=\,-4(1-cy)\,\,,
\eeq
which can be immediately integrated, namely:
\beq
D(y)\,=\,-4\bigg(\,y-{cy^2\over 2}\,\bigg)\,+\,{\rm constant}\,\,.
\label{D(y)}
\eeq
By performing a Legendre transform in (\ref{Leff}) we can obtain
the energy of the configuration:
\beq
E\,=\,{T_5 L^4\over 108}\,\,V_4\,\,\int  dy\,\,
{\cal H}\,\,,
\eeq
where ${\cal H}$ is given by:
\beq
{\cal H}\,=\,
(1-cy)\,H\,\sqrt{{r^2\over H^2}\,+\,6\,(r')^2\,-\,6\,(F_{x^0 y})^2}\,+\,
D(y)\,F_{x^0 y}\,\,.
\eeq
Moreover, the relation (\ref{D}) between $D(y)$ and $F_{x^0 y}$ can be inverted, with the
result:
\beq
F_{x^0 y}\,=\,{1\over 6}\,\,
{\sqrt{{r^2\over H^2}\,+\,6\,(r')^2}\over
\sqrt{{D^2\over 6}\,+\,(1-cy)^2\,H^2}}
\,\,D\,\,.
\label{invertedD}
\eeq
Using the relation (\ref{invertedD}) we can rewrite ${\cal H}$ as:
\beq
{\cal H}\,=\,\sqrt{{D^2\over 6}\,+\,(1-cy)^2\,H^2}\,\,\,
\sqrt{{r^2\over H^2}\,+\,6\,(r')^2}\,\,,
\eeq
where $D(y)$ is the function of the $y$ coordinate displayed in (\ref{D(y)}). 
The Euler-Lagrange equation derived from ${\cal H}$ is a second-order differential equation
for the function $r(y)$. This equation is rather involved and we will not
attempt to solve it
here. In a supersymmetric configuration one expects that there exists  a first-order
differential equation for $r(y)$ whose solution also solves the equations
of motion. This first-order equation has been found in
ref.\cite{severalbaryon} for the $AdS_5\times S^5$ background. We have
not been able to find such first-order equation in this 
$AdS_5\times Y^{p,q}$ case. A similar negative result was obtained in \cite{acr} for the 
$AdS_5\times T^{1,1}$ background. This result is an indication that this baryon vertex
configuration is not supersymmetric. Let us check explicitly  this fact by analyzing the
kappa symmetry condition. The expression of $\Gamma_{\kappa}$ when the worldvolume gauge
field is non-zero can be found in ref.\cite{swedes}. In our case
$\Gamma_{\kappa}\,\epsilon$ reduces to:
\beq
\Gamma_{\kappa}\,\epsilon\,=\,-{i\over \sqrt{-\det (\,g+F\,)}}\,\,\Bigg[\,
{r\over L}\,\Gamma_{x^0}\,\gamma_{y\beta\theta\phi\psi}\,\,\epsilon^*\,-\,
F_{x^0 y}\,\gamma_{\beta\theta\phi\psi}\,\,\epsilon\,\Bigg]\,\,.
\label{baryonkappa}
\eeq
The two terms on the right-hand side of (\ref{baryonkappa}) containing the antisymmetrized
products of gamma matrices can be written as:
\bear
&&\gamma_{y\beta\theta\phi\psi}\,\,\epsilon^*\,=\,
{L^5\over 108}\,(1-cy)\sin\theta\,\Big(\,\Gamma_{5}\,-\,
\sqrt{6}\,H\,{r'\over r}\,\Gamma_{r15}\,\Big)\,\epsilon^*\,\,,\rc\rc
&&\gamma_{\beta\theta\phi\psi}\,\,\epsilon\,=\,
-{L^4\over 18\sqrt{6}}\,(1-cy)\,H\,\sin\theta\,\Gamma_{15}\,\epsilon\,\,.
\eear
By using this result, we can write $\Gamma_{\kappa}\,\epsilon$ as:
\beq
\Gamma_{\kappa}\,\epsilon\,=\,
-{i\,L^4\,(1-cy) \over \sqrt{-\det (\,g+F\,)}}\sin\theta\Bigg[
{r\over 108}\,\Gamma_{x^0}\Gamma_5\,\epsilon^*+
{H\over 18\sqrt{6}}\,\Big(\,F_{x^0y}\,\Gamma_{15}\,\epsilon\,-\,
r'\,\Gamma_{x^0r15}\,\epsilon^*\,\Big)\Bigg]\,\,.
\label{baryongammakappa}
\eeq
In order to solve the $\Gamma_{\kappa}\,\epsilon\,=\,\epsilon$ equation we
shall impose, as in ref.\cite{Susybaryon}, an extra projection such that
the contributions of the worldvolume gauge field  $F_{x^0y}$ and of $r'$
in (\ref{baryongammakappa}) cancel  each other.  This can be achieved
by imposing that $\Gamma_{x^0r}\,\epsilon^*\,=\,\epsilon$  and that
$F_{x^0y}=r'$. Notice that the condition $\Gamma_{x^0r}\,\epsilon^*\,=\,\epsilon$
corresponds to having fundamental strings in the radial direction, as expected for a baryon
vertex configuration. Moreover, as the spinor $\epsilon$ has fixed ten-dimensional
chirality, this extra projection implies that 
$i\Gamma_{x^0}\Gamma_5\,\epsilon^*\,=\,-\epsilon$ which, in turn, is needed to satisfy the 
$\Gamma_{\kappa}\,\epsilon\,=\,\epsilon$ equation. However, the condition 
$\Gamma_{x^0r}\,\epsilon^*\,=\,\epsilon$  is incompatible with the conditions
(\ref{epsilon-project}) and, then, it cannot be imposed on the Killing
spinors. Thus, as in the analysis of \cite{acr}, we conclude from this
incompatibility argument (which is more general than the particular
ansatz we are adopting here) that the baryon vertex configuration breaks
completely the supersymmetry of the
$AdS_5\times Y^{p,q}$ background. 

\subsection{More spacetime filling D7-branes}
\label{MoreD7}

Let us adopt $\xi^{\mu}=(x^0, x^1, x^2,x^3,  y, \beta, \psi, r)$ as our set of worldvolume
coordinates for a D7-brane probe and let us consider a configuration with
$\theta=\theta(y,\beta)$ and $\phi=\phi(y,\beta)$.  In this case:
\beq
\Gamma_{\kappa}\,=\,-{i\over \sqrt{-g}}\,\,
{r^4\over L^4}\,\,\Gamma_{x^0x^1x^2x^3}\,\gamma_{y\beta\psi r}\,\,.
\eeq
Let us take $\epsilon=\epsilon_-$, where $\Gamma_*\epsilon_-=-\epsilon_-$(see eq.
(\ref{chiraladsspinor})). 
As $\gamma_r={L\over r}\,\Gamma_r$, we can write:
\beq
{r\over L^4}\,\,
\gamma_{y\beta\psi r}\,\epsilon_-\,=\,-\big[c_5\,+\,c_{135}\,\Gamma_{13}\,\big]\,
\epsilon_-\,\,,
\eeq
where the coefficients $c_5$ and $c_{135}$ are exactly those written in eq.
(\ref{expressionHEKcs}) for the D (doublet) three-cycles. The BPS
condition is just $c_{135}=0$, which leads to  the system of differential
equations (\ref{HEKBPS}). Thus, in this case the D7-brane extends 
infinitely in the radial direction and wraps a three-dimensional
submanifold of the 
$Y^{p,q}$ space  of the
type studied in subsection \ref{hekconst}. These embeddings preserve four supersymmetries.

\subsection{D7-branes wrapped on $Y^{p,q}$}
\label{D7Ypq}

Let us take a D7-brane which wraps the entire $Y^{p,q}$ space and is extended along two 
spatial directions. The set of worldvolume coordinates we will use in this case are
$\xi^{\mu}\,=\,(x^0,x^1,r,\theta,\phi,y,\beta,\psi)$ and we will assume that $x^2$ and $x^3$ are
constant. The matrix $\Gamma_{\kappa}$ in this case is:
\beq
\Gamma_{\kappa}\,=\,-{i\over \sqrt{-g}}\,\,
\gamma_{x^0x^1r\theta\phi y\beta\psi}\,\,.
\eeq
Acting on a spinor $\epsilon$ of the background one can prove that
\beq
\Gamma_{\kappa}\,\epsilon\,=\,i\Gamma_{x^0x^1r 5}\,\epsilon\,\,,
\eeq
which can be solved by a spinor $\epsilon_-=r^{{1\over 2}}\,\,e^{-{i\over 2}\psi}\,\eta_-$, with
 $\eta_-$ satisfying the additional projection $\Gamma_{x^0x^1r
5}\,\eta_-\,=\, -i\eta_-$. Thus this configuration preserves two
supersymmetries.

\setcounter{equation}{0}
\section{Summary and Conclusions}
\label{conclusions}
Let us briefly summarize the results of our investigation. Using
kappa symmetry as the central tool, we have  systematically studied
supersymmetric embeddings of branes in the $AdS_5\times Y^{p,q}$
geometry.  Our  study focused on three kinds of branes D3, D5 and D7.

{\it D3-branes: }This is the case that we studied most exhaustively. For 
D3-branes wrapping three-cycles in $Y^{p,q}$ we first reproduced 
all the results present in the literature. In particular, using
kappa symmetry, we obtained two kinds of  supersymmetric cycles:
localized at $y_1$ and $y_2$  \cite{ms} and  localized in the  round
$S^2$ \cite{sequiver,hek}. For these branes we found perfect agreement
with the field theory results.  Moreover, we also found a new class of
supersymmetric embeddings of D3-branes in this background. They do not
correspond to dibaryonic operators since the D3-brane does not wrap a
three-cycle.  The field theory interpretation of these new embeddings is
not completely clear to  us due to various issues with global properties.
We believe that they might be  a good starting point  to find candidates
for representatives of the  integer part of the third homology group of
$Y^{p,q}$, just like the analogous family of cycles found in
\cite{ba2,acr} were representative of the integer  part
$H_3(T^{1,1}, \ZZ)$. It would be important to understand these
wrapped D3-branes in terms of algebraic geometry as well as in terms of  
operators in the field theory dual, following the framework of
Ref.\cite{beasley} which, in the case of the conifold, emphasizes the use
of global homogeneous coordinates. It is worth stressing that such global 
homogeneous coordinates exist in any toric variety \cite{Cox} but the
relation to the field theory operators is less clear in $CY^{p,q}$. 
We analyzed the spectrum of excitations of a wrapped  D3-brane
describing an $SU(2)$-charged dibaryon and found perfect agreement with
the field theory expectations.  We considered other embeddings and found
that  a D3-brane wrapping a two-cycle in $Y^{p,q}$ is not a supersymmetric
state but, nevertheless, it is stable. In the field theory this 
configuration describes a fat string.

{\it D5-branes:}  The embedding that we paid the most attention to is a
D5-brane extended along a two-dimensional submanifold in $Y^{p,q}$ and 
having codimension one in $AdS_5$. In the field theory this is the kind
of brane that represents a  domain wall across which the  rank of the
gauge groups jumps. Alternatively, when we allow the D5-brane to extend
infinitely in the holographic direction, we  get a configuration
dual to a defect conformal field theory of the type analyzed in ref.
\cite{WFO} for the
$AdS_5\times S^5$ background.  We showed explicitly that such
configuration  preserves four supersymmetries and saturates the expected
energy bound. For this configuration we also considered  turning on a
worldvolume flux and found that it can be done in a supersymmetric way.
The flux in the  worldvolume of the brane provides a bending  of the
profile of the wall, analogously to what happens in $AdS_5\times S^5$
\cite{ST} and
$AdS_5\times T^{1,1}$
\cite{acr}. We  showed the
consistency of similar 
 embeddings in which the D5-brane wraps a different two-dimensional
submanifold in $Y^{p,q}$. We also considered D5-branes wrapping
three-cycles. This configuration  also looks like a domain wall in the
field theory dual but it does not have codimension one in $AdS_5$ and,
although it  cannot be supersymmetric,  it is stable. Finally, we
considered a D5-brane wrapping the whole
$Y^{p,q}$, which corresponds  to  the baryon vertex. We verified that,
as in the case of $T^{1,1}$, it is not a supersymmetric configuration.

{\it D7-branes:} With the aim of  introducing mesons in the
corresponding field theory,  we  considered spacetime filling D7-branes.
We explicitly showed that such configurations preserve  four
supersymmetries and found  the precise embedding in terms of the radial
coordinate. We  found an interpretation of the embedding equation in
terms  of complex coordinates.  We also analyzed other spacetime filling
D7-brane embeddings. Finally, we considered a D7-brane that  wraps
$Y^{p,q}$ and is codimension two in $AdS_5$.  This configuration looks,
from the field theory point of view, as a  string and preserves two
supersymmetries.

We would like to comment on various approximations made in the paper and point out some interesting open problems. 
We believe that our analysis, though carried out in the case of $Y^{p,q}$
manifolds, is readily adaptable to other  Sasaki-Einstein spaces. In
particular, the form of the spinor for $L^{a,b,c}$ is essentially the
same as in our case, namely  
$\epsilon^{-i\psi/2}\eta$,  where $\psi$ is the coordinate on the $U(1)$
fiber in the canonical presentation of   Sasaki-Einstein spaces as a
$U(1)$ bundle over  a K\"ahler-Einstein base,  and $\eta$ is a constant
spinor satisfying two projections generically written  as
$\Gamma_{12}\eta=-i\eta$ and $\Gamma_{34}\eta=i\eta$. Note that this
structure comes from the K\"ahler base and is  universal.

Part of our analysis of some branes could be made more precise. In particular, it would be interesting 
to understand the new family of supersymmetric embeddings of D3-branes
in terms of algebraic geometry as well  as in terms of
operators in the field theory. We did not present an analysis of the
spectrum of  excitations for all of the branes. In particular, we would
like to understand the excitations of the spacetime filling  D7-branes
and  the baryon vertex better. We hope that understanding the conformal
case will provide the basis for future analysis of  deformed theories
including the confining ones.  For example, as shown by \cite{LevO}
and \cite{Kuper}, the spacing of mass  eigenvalues for the mesons in
the confining case inherits properties of the related conformal theory.

We would like to point out that fully matching the spectrum of wrapped branes with field theory states 
is largely an ongoing problem. In particular, there are various
embeddings  that originate 
from branes of different dimensionality wrapping different cycles which
should be distinguishable from the field theory point of view. It
would be  interesting to understand to what extent the topological data
of the space determine the kind of supersymmetric branes that are 
allowed. Let us finish  with a wishful statement. We have found a large
spectrum of supersymmetric wrapped branes and also non-supersymmetric but
stable  branes. In analogy with the situation for strings in flat space
and orbifolds one wonders whether there is a sort of  holographic K-theory
which accounts for all the possible branes  in a given background.

We hope to return to these issues in the near future. 

\medskip
\section*{Acknowledgments}
\medskip
We are thankful to J. Liu and J. Shao for collaboration at early stages
of this project. We are grateful to Sergio Benvenuti, Ami Hanany,
Dario Martelli and Peter Ouyang for comments and to M. Mahato for various
checks. We would also like to thank David Mateos, Carlos N\'u\~nez and
Peter Ouyang for insightful comments on the manuscript.
F.C. and J.D.E. wish to thank the Theory Division at
CERN for hospitality while this work was being completed.
The research of L.A.P.Z. and D.V. was supported in part by Department of
Energy under grant DE-FG02-95ER40899 to the University of  Michigan and the National Science Foundation under Grant No. PHY99-07949 to the Kavli
Institute for Theoretical Physics. J.D.E. has been supported in part by
ANPCyT under grant PICT 2002 03-11624, and by the FCT grant
POCTI/FNU/38004/2001. The work of F.C., J.D.E. and A.V.R. was
supported in part by MCyT, FEDER and Xunta de Galicia under grant
FPA2005-00188 and by  the EC Commission under  
grants HPRN-CT-2002-00325 and MRTN-CT-2004-005104. 
Institutional support to the Centro de
Estudios Cient\'\i ficos (CECS) from Empresas CMPC is gratefully
acknowledged. CECS is a Millennium Science Institute and is funded
in part by grants from Fundaci\'on Andes and the Tinker Foundation.

\end{document}